# Empirically assessing the plausibility of unconfoundedness in observational studies

Fernando Pires Hartwig[1,2*], Kate Tilling[2,3], George Davey Smith[2,3]

[1]Postgraduate Program in Epidemiology, Federal University of Pelotas, Pelotas, Brazil.

[2]MRC Integrative Epidemiology Unit, University of Bristol, Bristol, United Kingdom.

[3]Population Health Sciences, University of Bristol, Bristol, United Kingdom.

*Corresponding author. Postgraduate Program in Epidemiology, Federal University of Pelotas, Pelotas (Brazil) 96020-220. Phone: 55 53 981126807. E-mail: fernandophartwig@gmail.com.

**Source of Funding:** The Medical Research Council (MRC) and the University of Bristol support the MRC Integrative Epidemiology Unit (MC_UU_00032/01 and MC_UU_00032/03). This work was supported by the Bristol Benjamin Meaker Follow-on Fund by the University of Bristol.

**Conflicts of interest:** None declared.

**Abstract**

The possibility of unmeasured confounding is one of the main limitations for causal inference from observational studies. There are different methods for (partially) empirically assessing the plausibility of unconfoundedness. However, most currently available methods require (at least partial) assumptions about the confounding structure, which may be difficult to know in practice. In this paper we describe a simple strategy for empirically assessing the plausibility of conditional unconfoundedness (i.e., whether the candidate adjustment set of covariates suffices for confounding adjustment) which does not require any explicit assumptions about the confounding structure, relying instead on assumptions related to temporal ordering between covariates, exposure and outcome (which can be guaranteed by design) and selection into the study. The proposed method essentially relies on testing the association between a subset of the covariates included in the adjustment set (those associated with the exposure, given all other covariates) and the outcome conditional on the remaining covariates and the exposure. We describe the assumptions underlying the method, provide proofs, use simulations to corroborate the theory and illustrate the method with an applied example assessing the causal effect of delivery mode and intelligence quotient measured in adulthood using data from the 1982 Pelotas (Brazil) birth cohort. We also discuss the implications of measurement error and some important limitations of the suggested approach.

## 1. Introduction

In observational studies, a major limitation for causal inference is the possibility of confounding.[1] There are several examples of apparently robust findings from observational studies being seriously questioned by subsequent randomized trials.[2-4] Of note, confounding is not the only possible explanation for different results between observational and intervention studies.[5] In the framework of directed acyclic graphs, confounding can be defined as the existence of open backdoor paths between the exposure (or risk factor or treatment) and outcome (or response) due to common causes.[1] However, open backdoor paths can also exist due to conditioning on a collider. Here we use the word "confounding" to mean the existence of any open backdoor path.

Although it is well-established that statistical methods cannot determine whether confounding exists without untestable assumptions,[1,6] methods for partial empirical verification of unconfoundedness can be useful.[7] Common examples include: covariate balance assessment following inverse probability weighting or propensity score matching (which assumes no unmeasured confounding);[8] comparing results from conventional covariate-adjusted analyses and instrumental variables (which assume that the instrument is valid, in which case they are robust to unmeasured confounding);[9] negative control exposures[10] or outcomes[11] (which assume that confounders between exposure and outcome are the same as the confounders between the negative control exposure and the outcome, or the confounders between the exposure and negative control outcome, respectively); and falsification tests for unconfoundedness[6,12] (which typically require relatively few assumptions, but cannot provide positive evidence in support of unconfoundedness). Another possible strategy uses conditioning on the exposure. For example, if there is a known valid instrument for the exposure, the association between the instrument and the outcome conditional on the exposure would be informative of the presence of confounding between the exposure and the outcome due to collider bias.[13] Importantly, all methods mentioned above require at least some assumptions about the confounding structure (i.e., unverifiable constraints on the causal relationships among exposure, outcome and measured and/or unmeasured factors). In practice, although some causal knowledge is often available, it can be difficult to justify the specific causal assumptions required.

We describe a procedure that exploits conditioning on the exposure to examine whether a candidate covariate set is sufficient to adjust for all confounding. The procedure is formulated non-parametrically so that it does not depend on specific modelling assumptions. We also discuss the role of parametric assumptions for implementing the method, particularly when covariate data are of high dimensionality. First, we describe the notation and assumptions (section 2). Then, we provide some intuition for the rationale underlying the procedure (section 3), which is presented and compared to previous methods[6,12] (section 4). Its application is illustrated in a real data example (section 5), and strengths and weaknesses discussed in section 6.

## 2. Notation and assumptions

Let $X$ and $Y$ respectively denote the exposure and the outcome. We assume that $X$ is time-fixed – i.e., a single measure at a specific point in time (e.g., birthweight rather than weight throughout childhood). Let $C$ denote the candidate adjustment set. We assume the data (i.e., $X$, $Y$ and $C$) are generated by some acyclic model, which may contain additional variables and, apart from the constraints imposed by assumptions 1-4 (mentioned below), is left unspecified. Therefore, it is not known whether $C$ is a sufficient adjustment set.

We assume:

- Assumption 1: $Y$ is not a cause of $X$. This is guaranteed in a longitudinal study where $X$ precedes $Y$.
- Assumption 2: None of the elements in $C$ are caused by $X$. This is guaranteed if all covariates precede $X$.
- Assumption 3: Selection into the analytical sample (i.e., the subset of the study participants eligible for analysis addressing the research question) is either not caused by $X$ or not conditionally associated with covariates in $C$ that fulfill the requirements to be used for the proposed test (described below) given all other covariates in $C$. That is, eligibility criteria, follow-up attrition, missing data etc. are not caused by (but may nevertheless be associated with) exposure or not conditionally associated with a subset of covariates (denoted as $Z$ in the examples below). This is guaranteed if $X$ is posterior to, and therefore does not cause, selection.
- Assumption 4: Faithfulness. Under this assumption, d-connection (i.e., existence of at least one open path between two variables) implies statistical dependence.
- Assumption 5: Correct model specification. Although assumptions 1-4 above are sufficient for implementing the proposed method (see below) non-parametrically, in practice $C$ is usually of high dimensionality, so that statistical adjustment requires modelling. Therefore, we also need to assume correct specification of two models: (a) the model for $X$ given $C$; and (b) the model for $Y$ given $X$ and $C$. As illustrated in the simulation study, model misspecification may distort the conclusions from our method even when assumptions 1-4 hold.

## 3. Intuition for the proposed method

For some intuitive understanding, consider Figure 1a. Suppose that $C = D$ – i.e., out of the set of measured covariates $\{Z, D\}$, the adjustment set $C$ only includes the single covariate $D$. Now consider two possibilities within this scenario: (i) $U = \emptyset$ (i.e., there are no unmeasured confounders) and (ii) $U \neq \emptyset$ (i.e., there are unmeasured confounders). The goal is to use measured variables ($Z$, $D$, $X$, and $Y$) to choose between (i) or (ii). One possibility is to test the association between $Z$ and $Y$ conditional on $X$ and $C$. In this case, the only open path between $Z$ and $Y$ would be the $Z \rightarrow \boxed{X} \leftarrow U \rightarrow Y$ (this is because the path $Z \rightarrow \boxed{X} \rightarrow Y$ is blocked by conditioning on $X$ and the path $Z \rightarrow \boxed{X} \leftarrow C \rightarrow Y$ is blocked by conditioning on $C$). Therefore, $Z$ and $Y$ would be conditionally d-connected if $U \neq \emptyset$ and d-separated if $U = \emptyset$. By Assumption 4, conditional associations between $Z$ and $Y$ could therefore be used as a test to check if $C$ is a valid adjustment set in this situation. Note that the same conclusions applies if $Z$ is not a cause of $X$ (Figure 1b).

This simple example illustrates the role of Assumptions 1-4 (Figure 1c-e contains examples of violations of Assumptions 1-3). Notice that $Z$, a variable associated with $X$ given $C = D$, was used to assess if $C$ is a valid adjustment set. By Assumptions 2 and 3, paths of the form $X \rightarrow Z$ and $X \rightarrow \boxed{V} \leftarrow Z$ cannot occur. Therefore, any dependency between $Z$ and $X$ implies that there are one or more open (given $C$) paths between $Z$ and $X$ with an arrow pointing at $X$. This in turn implies that there are paths between $Z$ and $U$ (as well as all other causes of $X$) where $X$ is a collider. Therefore, by conditioning on $X$, any path between $Z$ and $Y$ of the form $Z \rightarrow \boxed{X} \rightarrow Y$ (or $Z \leftarrow V \rightarrow \boxed{X} \rightarrow Y$) is blocked, and any path of path of the form $Z \rightarrow \boxed{X} \leftarrow U \rightarrow Y$ is open.

Therefore, if there are unmeasured confounders, $Z$ and $Y$ will be d-connected given $X$. Given Assumption 4, this implies that the conditional (on $C$ and $X$) association (or lack thereof) between $Z$ and $Y$ provides information about unmeasured confounders. Assumption 1 is required because, if $Y \rightarrow X$, then causes of $Y$ would be associated with $Y$ given $D$ even in the absence of confounding. Finally, Assumption 4 is required to ensure that d-connection implies association. This is important to avoid scenarios, for example, where there are unmeasured confounders such that $Z$ and $Y$ are d-connected given $X$, but $Z$ and $Y$ are not associated due to some form of cancellation (e.g., due to multiple open paths of opposing directions).

Notice, however, that Assumptions 1-4 are compatible with multiple causal structures. For example, in Figure 1f, all the arguments above apply and testing the association between $Z$ and $Y$ conditional on $X$ can be used to test if $C$ is a valid adjustment set: if $C = D$, then $Z$ and $Y$ are unassociated conditional on $X$ and $A$; otherwise, $Z$ and $Y$ would be associated via $Z \rightarrow D \rightarrow Y$ and $Z \rightarrow \boxed{X} \leftarrow D \rightarrow Y$. This is also true in Figure 1g; indeed, in this case $C = D$ would be clearly an invalid adjustment set since $Z$ would itself be a confounder.

In all examples discussed so far, $Z$ can be used to assess whether $C = D$ is a valid adjustment set. However, there are also examples where this fails. Consider the situation illustrated in Figure 1h. Notice that $D$ is a valid adjustment set because the only other backdoor path between $X$ and $Y$ is blocked at $Z$. However, $Z$ and $Y$ would be d-connected given $X$ and $D$. So, even though $D$ is a valid adjustment set, $Z$ and $Y$ would be conditionally associated, unlike the examples discussed above.

The above limitations can be solved in at least two ways. The simplest one is to include the additional assumption that $Z$ causes $X$. An alternative solution, which does not require additional assumptions, requires modifying the perspective presented above. Notice that, so far, all examples discussed assessing the validity of $C$ for confounding adjustment using an element outside the adjustment set, which we denoted by $Z$. However, now consider the case where $Z \in C$ (i.e., $Z$ is one of the covariates in the candidate adjustment). That is, the idea is now to use one or more elements in the adjustment set (instead of elements outside of the adjustment set) to assess the validity of the adjustment set (which now includes $Z$).

It is clear that $C = \{D \cup Z\}$ is a valid, but not minimal, adjustment set in Figure 1a,b,f,g. This is because $Z$ can only be associated with $Y$ via $X$ (after conditioning on $D$) and $D$ itself is a valid adjustment set. Moreover, the problem illustrated in Figure 1h no longer exists, since in this case $C = \{D \cup Z\}$ is an invalid adjustment set (because conditioning on $Z$ opens a backdoor path between $X$ and $Y$) and $Z$ and $Y$ are d-connected given $X$ and $D$. The downside would be in situations like Figure 1g, where $C$ is a valid adjustment set but $Z$ is nevertheless associated with $Y$ given $X$ and $D$. This indicates that, if $C = \{D \cup Z\}$, then conditional association between $Z$ and $Y$ does not necessarily imply that $C$ is an invalid adjustment set, but lack of conditional association implies $C$ is a valid adjustment set.

**4. Assessing unconfoundedness**

Our main result is the following procedure for empirically assessing the plausibility of unconfoundedness. Suppose that there is at least one covariate (denoted by $Z$) in the covariate set $C$ such that: (i) $Z$ is associated with (i.e., not independent of) $X$ given all other covariates (denoted by $D$, so $C = \{D \cup Z\}$); and (ii) $Z$ and $Y$ are conditionally independent given $X$ and $D$. In this case, $C$ is a valid adjustment set. Conversely, if none of the covariates that satisfy condition (i) also satisfy condition (ii), $C$ is either an invalid adjustment set or a minimal valid adjustment set in the sense that $Z$ and $D$ are, when considered separately, invalid adjustment sets and $C$ is a valid adjustment set (of note, this ambiguity can be eliminated if $Z$ is assumed to be exogenous – i.e., has no common cause with any other variable also included in the data-generating acyclic model – and $Z$ is not part of the covariate set being tested; see supplementary section 1 for details). Therefore, the result is only conclusive if some $Z$ satisfies both (i) and (ii). In the supplement, the approach is formalized and proven (section 1), and applied for time-varying exposures (section 2).

In supplementary section 3, we discuss the implications of measurement error in $X$, $Y$ and $C$. Briefly, $C$ can be mismeasured because we are not interpreting $C$ as if it were identical to its corresponding theoretical set of perfectly measured variables. This is desirable because any actual set of covariates is likely imperfectly measured. Moreover, the procedure will assess whether $C$ is sufficient for bias adjustment regarding the effect of the measured version of $X$ on the measured version of $Y$. As discussed in the supplement, the measurement error structure influences how results obtained using mismeasured $X$ and/or $Y$ relate to results that would have been obtained in the absence of measurement error. Whether measurement error correction techniques – which often require strong assumptions on the error structure and magnitude[14,15] – could be used in combination with the proposed procedure requires further methodological research.

There are previous tests with important similarities to ours (see supplementary section 4 for a detailed comparison). One of them was proposed by Pearl (Theorem 6.4.4).[6] The main differences are that our test requires faithfulness (Pearl's does not) and, if its assumptions are satisfied, can correctly classify $C$ as a valid adjustment set, but not as an invalid adjustment set (because $C$ could be minimal in the sense described in section 3). Pearl's test requires that the variable used in the test (analogous to $Z$ in our context) is not a barren proxy (i.e., a variable associated with both the exposure and the outcome without causally affecting either; our test does not need this assumption) and, if its assumptions are satisfied, is a falsification test. Of note, supplementing it with faithfulness does not enable the test to correctly identify $X$ and $Y$ as unconfounded, while supplementing our test with the assumption that $Z$ is not a barren proxy allows for conclusive results regarding $D$ being a valid adjustment set. Entner and colleagues[12] describe an automated covariate selection procedure which can be seen as a generalization of our approach. However, our method has important advantages in relation to its practical application: much lower computational requirements, lower risk of near faithfulness violations by not assessing all possible subsets of $C$, and a clearer separation between covariate selection and causal effect estimation.

## 5. Applied example

In addition to simulations performed to corroborate the theoretical results (see supplementary section 5 for a detailed description of the methodology and results), we applied the proposed method to data from the 1982 Pelotas (Brazil) Birth Cohort for illustration. Details on eligibility criteria, follow-up history and data availability are given elsewhere.[16,17] The exposure ($X$) is delivery mode (exposed: c-section; unexposed: vaginal) measured at the perinatal interview, and the outcome ($Y$) is intelligence quotient (IQ) measured at 30 years. Given the strong social patterning in delivery mode[18] and cognitive development[19] in Pelotas, socioeconomic factors are likely the most relevant confounders. We therefore selected three maternal covariates related to this concept: self-reported skin color (white/other), schooling (numeric, expressing complete years of education) and socioeconomic position (categorized in quintiles)[20] at birth. We also included in the covariate set maternal smoking (categorized in 0, 1-14 cigarettes and ≥15 cigarettes over the entire pregnancy period), age and pre-pregnancy body mass index (BMI) (both numeric), parity (categorized as 0, 1, 2, 3 or ≥4), as well as individual's sex. These eight variables are the covariate set ($C$) we are examining.

Our analyses included 3087 individuals (out of 5914 recruited at baseline) with complete data for exposure, outcome and covariates. The outcome was standardized to have zero mean and unit standard deviation (SD). Without covariate adjustment, c-section was associated with 0.29 (95% CI: 0.22; 0.37) higher SD units of IQ.

We explored different strategies for modeling the association of (1) exposure with covariates, and (2) outcome with exposure and covariates. For model 1, we selected a logistic regression model with a simple specification (i.e., not including transformation or higher order terms). For model 2, we selected a linear regression model with a simple specification (details on the model selection process can be found in supplementary section 6 and supplementary table 4). Based on model 2, c-section was associated with 0.04 (95% CI: -0.03; 0.11) higher SD units of IQ.

We now assess the plausibility that the set of eight covariates mentioned above is sufficient for confounding adjustment. First, we used model 1 to check if any covariate is associated with the exposure conditional on all other covariates (i.e., if any covariate satisfies condition (i)). Second, we used model 2 to check if any covariate is not associated with the outcome conditional on the exposure and all other covariates (i.e., if any covariate satisfies condition (ii)). Under assumptions 1-5, if at least one covariate satisfies both (i) and (ii), then our covariate set is sufficient for confounding adjustment.

As shown in Table 1, all covariates except maternal skin color, maternal smoking and sex were associated with the exposure conditioning on all other covariates in model 1. Since these three covariates do not satisfy condition (i), they cannot be used for assessing confoundedness. Among the remaining five covariates, only for maternal pre-pregnancy BMI there was only very weak evidence against the null hypothesis of no association with the outcome (given the exposure and all other covariates) in model 2. Therefore, this covariate satisfies both (i) and (ii), which would indicate that the entire covariate set is a valid adjustment set (as described in supplementary section 1, we would also conclude that a smaller valid adjustment set could be obtained by removing maternal skin colour, maternal smoking and sex from the covariate set). Model 3 is similar to model 2, except that socioeconomic variables are excluded from the covariate set. Since these are likely the strongest confounders, if faithfulness holds and there is sufficient statistical power, then it would be expected that all covariates associated with the exposure in model 1 are also associated with the outcome in model 3. Indeed, this was the case for maternal BMI. In a fourth model including only delivery mode and maternal BMI as predictors of IQ, the evidence for an association between maternal BMI and IQ was slightly stronger than in Model 3. Therefore, maternal BMI seems to be able to detect residual confounding introduced by removing important variables from the covariate set.

To further assess whether the covariate set is sufficient for confounding adjustment, we used school/university type (none, public or private, assessed at the 2004-2005 follow-up visit) as a negative control outcome.[11] This variable is strongly socially patterned in Pelotas (so it is affected by the arguably most important confounders) and is unlikely to be caused by the exposure (delivery mode). The analysis included the 3631 cohort participants with non-missing data for delivery mode, covariates and the negative control. The covariate-unadjusted prevalence ratio of c-section was 1.26 (95% CI: 1.11; 1.42) and 1.49 (95% CI: 1.31; 1.69) comparing public vs. none and private vs. none, respectively. The corresponding estimates after covariate adjustment were 1.02 (95% CI: 0.93; 1.13) and 1.07 (95% CI: 0.96; 1.20). This is in accordance with the notion that, while unadjusted estimates are likely biased, the covariate set is sufficient for confounding adjustment.

Since delivery mode and maternal pre-pregnancy BMI both precede selection into the analytical sample, there is no guarantee that Assumption 3 is satisfied. We therefore assessed the plausibility of this assumption by testing the association between delivery mode and inclusion in the analytical sample. In analysis without covariate adjustment (n=5914, since delivery mode had no missing data), the prevalence ratio of being included in the analytical sample was approximately 0.95 (95% CI: 0.90; 1.01) comparing exposed vs. unexposed groups. In analysis with covariate adjustment (n=4970, after restricting to those with non-missing covariate data), the prevalence ratio was 1.02 (95% CI: 0.95; 1.09). So, there is no strong evidence for selection post enrollment being influenced by delivery mode, which corroborates Assumption 3.

In addition to Assumption 3, the fact that selection is posterior to delivery mode could introduce collider bias if delivery mode influences selection (e.g., by influencing survival). Importantly, such bias would not be detectable by the proposed approach, which can only detect bias due to open backdoor paths (i.e., paths pointing at, rather than from, the treatment variable). Therefore, interpreting the results above as valid causal effect estimates for the subpopulation of those who had complete data on treatment, outcome and covariates requires assumptions about selection bias even if unconfoundedness is assumed to be true. As mentioned above, there was no strong indication that delivery mode is associated with selection, which provides at least some corroboration to the assumption that delivery mode does not influence selection in the empirical example.

Finally, we estimated the power of the proposed method to detect different levels of residual confounding. In simulations (supplementary section 5), the capacity of the method of detecting unmeasured confounding was strongly dependent on the underlying data generating process, thus making it difficult to derive analytical power calculation procedures that are generally applicable. To allow the structure of the dataset at hand to inform the power calculations, these were performed via plasmode simulations to preserve the full covariate structure observed in the data, as well as the associational pattern between the exposure and covariates.[21] Before doing the power calculations, we performed additional simulations aimed at assessing the performance of this plasmode power calculation in this dataset (full description of power calculations and additional simulations in supplementary section 7). In short, the power calculation approach was somewhat conservative (i.e., the estimated capacity to detect confounding was smaller than the true performance of the method).

The maximum confounding we simulated biases the linear regression coefficient of IQ on delivery mode by 0.4 SD units. Nevertheless, the power to detect even such substantial unmeasured confounding was low: the estimated probability of observing one or more covariates, among those that satisfied condition (i), not satisfying condition (ii) was 41% or more (depending on some inputs to the calculation, explained in detail in supplementary section 7 and shown in supplementary table 5). This indicates that, in the applied example, having only a single covariate satisfying both (i) and (ii) is not unexpected even if there were a strong unmeasured confounder, thus not providing strong support for unconfoundedness. Therefore, our analysis may be underpowered, which could lead to lack of strong statistical evidence for association between maternal pre-pregnancy BMI and the outcome in model 2 even if the covariate set is not sufficient for confounding adjustment.

## 6. Discussion

We propose a strategy for empirically assessing the plausibility of unconfoundedness. We considered several relevant aspects of an epidemiological study, including temporal relationships between variables and the possibility of selection bias and measurement error. In addition to theoretical proofs corroborated by simulations, our applied example illustrates how to implement the method in practice and the importance of power calculations.

Since our approach cannot differentiate whether $C$ is invalid or minimal, it is more applicable to situations where $C$ is unlikely to be minimal – e.g., $C$ if was conservatively selected and therefore likely contains redundant covariates (in the sense that adjusting for only one out of two or more would be sufficient to block a given biasing path). This is frequent in applied research because there is often considerable uncertainty in the causal assumptions needed for covariate selection, often leading to selecting a large number of pre-exposure covariates with no clear causal justification. Moreover, due to measurement error in the covariates or when attempting to adjust for latent constructs, researchers often select covariates known to be closely related and therefore possibly redundant (e.g., select both income and assets as covariates for adjusting for socioeconomic position). In the case where all covariates are conditionally associated with the outcome, bias amplification (i.e., larger bias due to conditioning on an instrument) will be a possibility. This is because, in this situation, the adjustment set may be invalid (more specifically, it is either valid or minimally valid, as described in section 4) and some covariates could be conditional instruments.[22,23] If there is suspicion that the covariate set both is invalid and contains conditional instruments, sensitivity analysis could be conducted by removing the possible conditional instruments from the adjustment set. If some covariates are known instruments, then they should be removed from the covariate set and could even be used for assessing unconfoundedness as described in section 4.

Importantly, our method is not primarily aimed at covariate selection, but for testing if a selected covariate set is sufficient for confounding adjustment. In the supplement (section 8), we discuss how covariate selection strategies[12,24-26] could be used in combination with our method. One possible misuse of the proposed approach is reporting that a covariate set was defined *a priori* (e.g., based on causal assumptions) when it was in fact selected by an automated search over a large covariate space until finding a covariate set that satisfies the Theorem's requirements and yields desired results. The risk of this bad practice could be mitigated by pre-registration of analysis protocols, which has been promoted for avoiding "hacking" (intentional selection and report of data or statistical analysis that produce the desired results) or "harking" (when allegedly *a priori* hypotheses are in fact written after the results are known) procedures.[27,28]

Violations or near-violations of faithfulness and low statistical power are difficult (if at all possible) to distinguish in finite samples, and their consequences for the proposed method are similar. It seems plausible that, the more covariates conditionally associated with the exposure and not associated with the outcome, the lower the plausibility that the results are due to faithfulness violations (but could result from low power). Another way to partially assess statistical power and the plausibility of faithfulness jointly is as follows. Suppose $Z$ satisfies both conditions (i) and (ii) of the Theorem. If $Z$ also satisfies both conditions when using a subset of $C$ lacking known important confounders, then it would suggest that bias using the complete $C$ was not detected due to low power and/or faithfulness violations. This was illustrated in the applied example by comparing models 2-4. Yet another possible way to assess faithfulness is via cross-cohort comparisons with different confounding structures since violations of faithfulness are unlikely to occur in the same way between cohorts. So, if the same $Z$ is not associated with $Y$ given $X$ and $D$ in both cohorts (assuming $Z$ satisfies condition (i) in both), this would be supportive of $C$ being a valid adjustment set rather than violations of faithfulness.

The assumption that selection into the analytical sample is either not caused by $X$ or not associated with $Z$ given $D$ (assumption 3) is quite strong in longitudinal studies. Selection bias adjustment strategies such as inverse probability weighting[1,29,30] or multiple imputation could be used to make this assumption more plausible.[29,31,32] The applied example illustrates that, when $X$ has complete data, the plausibility of assumption 3 can be assessed via the association between $X$ and a missingness indicator.

The proposed approach relies on inferring statistical dependence or independence. In finite samples, such inferences are respectively prone to type I and type II error. One way to mitigate this problem is to use a stringent cutoff for inferring dependence (for example, a low p-value threshold when assessing condition (i) described in section 3) and a relaxed cutoff for inferring independence (for example, a large p-value threshold when assessing condition (ii) described in section 3).[12] This would increase the method's sensitivity at the expense of its specificity, as illustrated in the simulation study. There are several different approaches for variable selection, and exploring their applicability to this method is an important area for further research, especially if the covariates are of high dimensionality.

In large samples, the statistical power of the procedure may be high even if the bias is small, while in small studies the procedure might be underpowered to detect even considerable bias. Moreover, low power could also happen due to other factors – e.g., high collinearity between covariates or model misspecification. We therefore recommend power analysis (illustrated in section 5) when applying the procedure in practice. Our approach serves for bias detection, but not for bias quantification, which would require additional modelling assumptions. Indeed, there are examples of bias quantification that also exploit conditioning on the exposure,[13] which can be useful if the modelling assumptions are considered plausible. Our method should be viewed as one tool that can be triangulated with other strategies for strengthening causal inference.[33] Hopefully our results will contribute to improve robustness of findings from observational studies relying on statistical covariate adjustment, and therefore to the overall triangulation process.

**Table 1. Point estimates, confidence intervals and P-values describing the association of covariates with delivery mode (odds ratio) and with intelligence quotient (IQ) at age 30 years (linear regression coefficients) in the 1982 Pelotas Birth Cohort (n=3087[a]).**

| Independent variable | Model 1 - delivery | Model 2 - IQ | Model 3 - IQ |
|---|---|---|---|
| **Delivery mode** | - | P=0.221 | P=$1.7\times10^{-5}$ |
| Vaginal | - | 0 (Reference) | 0 (Reference) |
| C-section | - | 0.04 (-0.03; 0.11) | 0.16 (0.09; 0.24) |
| **Maternal age** | P=$2.8\times10^{-8}$ | P=$1.6\times10^{-8}$ | P=$4.0\times10^{-27}$ |
| Per-year | 1.05 (1.03; 1.07) | 0.02 (0.01; 0.03) | 0.04 (0.03; 0.04) |
| **Maternal skin color** | P=0.157 | P=$9.4\times10^{-11}$ | P=$3.6\times10^{-26}$ |
| White | 1 (Reference) | 0 (Reference) | 0 (Reference) |
| Other | 0.84 (0.65; 1.07) | -0.30 (-0.39; -0.21) | -0.49 (-0.58; -0.40) |
| **SEP at birth** | P=0.0002 | P=$1.6\times10^{-13}$ | - |
| Quintile 1 | 1 (Reference) | 0 (Reference) | - |
| Quintile 2 | 1.43 (1.06; 1.94) | 0.16 (0.06; 0.27) | - |
| Quintile 3 | 0.95 (0.69; 1.30) | 0.17 (0.06; 0.27) | - |
| Quintile 4 | 1.47 (1.08; 2.02) | 0.41 (0.30; 0.52) | - |
| Quintile 5 | 1.77 (1.20; 2.63) | 0.51 (0.36; 0.66) | - |
| **Maternal schooling** | P=0.0009 | P=$3.5\times10^{-23}$ | - |
| Per-year | 1.05 (1.02; 1.08) | 0.06 (0.05; 0.07) | - |
| **Parity** | P=$3.9\times10^{-10}$ | P=$2.4\times10^{-15}$ | P=$1.8\times10^{-43}$ |
| 0 | 1 (Reference) | 0 (Reference) | 0 (Reference) |
| 1 | 0.58 (0.47; 0.72) | -0.21 (-0.29; -0.13) | -0.26 (-0.34; -0.17) |
| 2 | 0.69 (0.53; 0.89) | -0.23 (-0.34; -0.13) | -0.37 (-0.48; -0.26) |
| 3 | 0.42 (0.28; 0.63) | -0.32 (-0.46; -0.18) | -0.59 (-0.74; -0.44) |
| ≥4 | 0.31 (0.21; 0.47) | -0.59 (-0.73; -0.45) | -1.04 (-1.19; -0.90) |
| **Maternal BMI** | P=$3.1\times10^{-10}$ | P=0.9690 | P=0.0237 |
| Per-kg/m² | 1.08 (1.05; 1.10) | 0.00 (-0.01; 0.01) | -0.01 (-0.02; 0.00) |
| **Maternal smoking** | P=0.721 | P=0.0040 | P=$2.5\times10^{-5}$ |
| 0 cigarettes | 1 (Reference) | 0 (Reference) | 0 (Reference) |
| 1-14 cigarettes | 0.93 (0.76; 1.13) | -0.08 (-0.15; -0.01) | -0.12 (-0.20; -0.04) |
| ≥15 cigarettes | 1.02 (0.74; 1.39) | -0.17 (-0.28; -0.05) | -0.25 (-0.37; -0.13) |
| **Sex** | P=0.1560 | P=0.0139 | P=0.1090 |
| Male | 1 (Reference) | 0 (Reference) | 0 (Reference) |
| Female | 0.89 (0.75; 1.05) | -0.08 (-0.14; -0.02) | -0.05 (-0.12; 0.01) |

Model 1: logistic regression of delivery more on all covariates simultaneously. Covariates associated with delivery mode satisfy condition (i) described in section 3.

Model 2: linear regression of IQ on deliver mode and all covariates simultaneously. Covariates not associated with IQ satisfy condition (ii) described in section 3.

Model 3: Same as model 2 but excluding socioeconomic covariates (SEP at birth and maternal schooling). This is a sensitivity analysis where likely important confounders were excluded to explore the capacity of the approach to detect unmeasured confounding (see section 5 for details).

SEP: Socioeconomic position. BMI: body mass index.

[a]All analyses were restricted to those with non-missing data for delivery mode, IQ and all covariates.

**Figure 1. Directed acyclic graphs illustrating the proposed test for unconfoundedness. In all scenarios, $X$, $Y$, $S$ and $\{Z, D\}$ respectively represent the exposure, outcome, selection into the study and measured covariates. $K$, $L$, $U$ and $W$ represent unmeasured variables.**

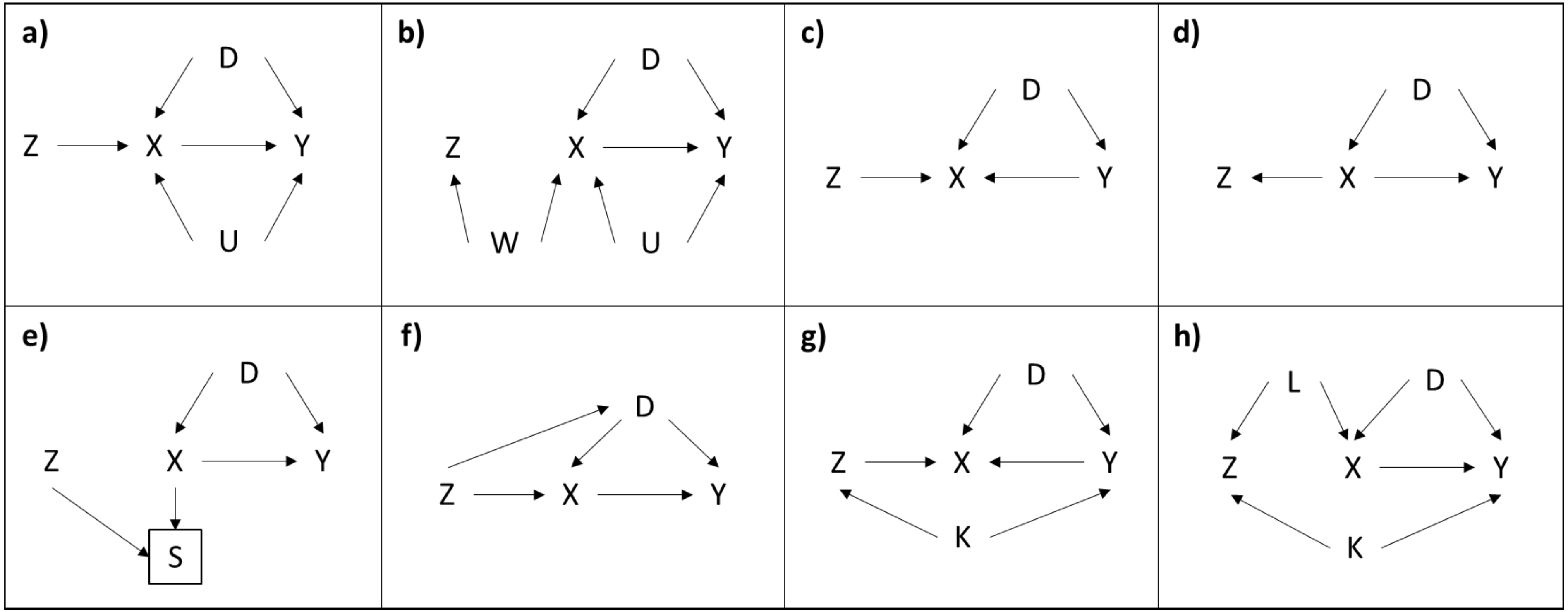

a) Assumptions 1-3 satisfied, and $Z$ causes $X$.
b) Assumptions 1-3 satisfied, and $Z$ does not cause $X$.
c) Assumption 1 violated.
e) Assumption 2 violated.
f) Assumption 3 violated.
g) Assumptions 1-3 satisfied, and $Z$ and $Y$ are conditionally (on $C$) d-connected only through $X$.
h) Assumptions 1-3 satisfied, $Z$ and $Y$ are d-connected even upon conditioning on $X$ and $C$, and $Z$ is a non-collider on a backdoor path between $X$ and $Y$.
i) Assumptions 1-3 satisfied, $Z$ and $Y$ are d-connected even upon conditioning on $X$ and $C$, and $Z$ is a collider on a backdoor path between $X$ and $Y$.

**Supplementary material for "Empirically assessing the plausibility of unconfoundedness in observational studies" by Hartwig FP, Tilling K and Davey Smith G**

**Summary**

## 1. Theorem

We now state a formal version of our main result, which is described rather informally in section 3 in the main text.

Theorem: Let $Z = \{Z_j\}_{j=1}^{J}$ denote a non-empty subset of $C$ which satisfies the following condition: $Z_j$ and $X$ are dependent conditional on $C_j^* = \{C \cap Z_j^C\}$ (where $Z_j^C$ is the complement of $Z_j$) for all $j = 1, \dots, J$. If $Z_j$ and $Y$ are dependent conditional on $\{C_j^* \cup X\}$ for all $j = 1, \dots, J$, then $C$ is either an invalid or a minimally sufficient adjustment set in the following sense: for all $j = 1, \dots, J$, given adjustment for $C_j^*$, there will be no confounding if and only if $Z_j$ is additionally adjusted for. Otherwise – that is, if $Z_j$ and $Y$ are independent conditional on $\{C_j^* \cup X\}$ for at least one $j \in \{1, \dots, J\}$, then $C$ is a valid (but non-minimal) adjustment set, and $C_j^*$ is also a valid (possibly, but not necessarily, minimal) adjustment set.

To prove the Theorem, we first define the following lemmas:

Lemma 1: Dependence between $Z_j$ and $X$ conditional on $C_j^*$ implies there is at least one open (given $C_j^*$) path between $Z_j$ and $X$ that is directed into $X$ (i.e., of the form $Z_j \rightarrow \cdots \rightarrow X$ or $Z_j \leftarrow \cdots \rightarrow X$).

Proof of Lemma 1: the existence of at least one open path follows from the notion that d-connection is required for dependence. Paths of the form $Z_j \rightarrow \cdots \leftarrow X$ or $Z_j \leftarrow \cdots \leftarrow X$. can only be conditionally open if they contain a collider than is being conditioned on (this is true even for the second path, because $Z_j$ is not caused by $X$ by Assumption 3). However, neither $C$ nor $S$ are caused by $X$ (Assumptions 3 and 4, respectively), so it is not possible that these paths contain $C$ or $S$ as colliders. Since no other variable is being conditioned on, these paths are closed. Therefore, it must be the case that there is at least one conditionally open path between $Z_j$ and $X$ that is directed into $X$.

Lemma 2: $Z_j$ and $Y$ can only be d-connected conditional on $\{C_j^* \cup X\}$ through paths that do not include $X$ as a non-collider (i.e., paths that either include $X$ as a collider or do not include $X$ at all).

Proof of Lemma 2: from Assumption 2, $X$ is measured without error. So, conditioning on $X$ blocks all paths where $X$ is a non-collider.

Corollary 1: $Z_j$ and $Y$ can only be d-connected conditional on $\{C_j^* \cup X\}$ through paths of the following form:

- 1: $Z_j \rightarrow \underset{....}{(a1)} \rightarrow \boxed{X} \leftarrow \underset{....}{(b1)} \rightarrow Y$.
- 2: $Z_j \rightarrow \underset{....}{(a1)} \rightarrow \boxed{X} \leftarrow \underset{....}{(b2)} \leftarrow Y$.
- 3: $Z_j \leftarrow \underset{....}{(a2)} \rightarrow \boxed{X} \leftarrow \underset{....}{(b1)} \rightarrow Y$.
- 4: $Z_j \leftarrow \underset{....}{(a2)} \rightarrow \boxed{X} \leftarrow \underset{....}{(b2)} \leftarrow Y$.
- 5: $Y \leftarrow \underset{....}{(c1)} \leftarrow Z_j \rightarrow \underset{....}{(a1)} \rightarrow \boxed{X}$
- 6: $Y \leftarrow \underset{....}{(c1)} \leftarrow Z_j \leftarrow \underset{....}{(a2)} \rightarrow \boxed{X}$
- 7: $Y \leftarrow \underset{....}{(c2)} \rightarrow Z_j \rightarrow \underset{....}{(a1)} \rightarrow \boxed{X}$
- 8: $Y \leftarrow \underset{....}{(c2)} \rightarrow Z_j \leftarrow \underset{....}{(a2)} \rightarrow \boxed{X}$
- 9: $Y \rightarrow \underset{....}{(c3)} \rightarrow Z_j \rightarrow \underset{....}{(a1)} \rightarrow \boxed{X}$

- 10: $Y \rightarrow \overset{(c3)}{\dots\dots} \rightarrow Z_j \leftarrow \overset{(a2)}{\dots\dots} \rightarrow \boxed{X}$
- 11: $Y \rightarrow \overset{(c4)}{\dots\dots} \leftarrow Z_j \rightarrow \overset{(a1)}{\dots\dots} \rightarrow \boxed{X}$
- 12: $Y \rightarrow \overset{(c4)}{\dots\dots} \leftarrow Z_j \leftarrow \overset{(a2)}{\dots\dots} \rightarrow \boxed{X}$

Corollary 2: $Z_j$ and $X$ may be d-connected conditional on $C_j^*$ through paths including $Y$:

- 13: $Z_j \rightarrow \overset{(c1)}{\dots\dots} \rightarrow Y \leftarrow \overset{(b1)}{\dots\dots} \rightarrow X$
- 14: $Z_j \rightarrow \overset{(c1)}{\dots\dots} \rightarrow Y \rightarrow \overset{(b2)}{\dots\dots} \rightarrow X$
- 15: $Z_j \leftarrow \overset{(c2)}{\dots\dots} \rightarrow Y \leftarrow \overset{(b1)}{\dots\dots} \rightarrow X$
- 16: $Z_j \leftarrow \overset{(c2)}{\dots\dots} \rightarrow Y \rightarrow \overset{(b2)}{\dots\dots} \rightarrow X$
- 17: $Z_j \leftarrow \overset{(c3)}{\dots\dots} \leftarrow Y \leftarrow \overset{(b1)}{\dots\dots} \rightarrow X$
- 18: $Z_j \leftarrow \overset{(c3)}{\dots\dots} \leftarrow Y \rightarrow \overset{(b2)}{\dots\dots} \rightarrow X$
- 19: $Z_j \rightarrow \overset{(c4)}{\dots\dots} \leftarrow Y \leftarrow \overset{(b1)}{\dots\dots} \rightarrow X$
- 20: $Z_j \rightarrow \overset{(c4)}{\dots\dots} \leftarrow Y \rightarrow \overset{(b2)}{\dots\dots} \rightarrow X$

Paths 13-20 can be obtained by combining segments (b) and (c) above. The fact that paths 13-20 are directed into $X$ follows from Lemma 1. Nevertheless, for clarity, we note that paths directed from $X$ are not allowed because they must be causal paths (i.e., paths that explain the causal effect of $X$ on $Y$), because paths of other forms would only be open if they contained $C_j^*$ or $S$ as a collider, which is not the case given Assumptions 3 and 4. However, for the paths to be open at $Y$, either $Y$ is a collider (as in paths 13 or 15), but it causes either $C$ or $S$, which is not allowed if $X$ causes $Y$ (but allowed if $X$ does not cause $Y$) because this would imply that $X$ causes such nodes; or $Y$ is a non-collider and the path between $Y$ and $Z_j$ is open, which can only happen if $Y$ causes $Z_j$ (which would imply $X$ causes $Z_j$, which is not the case given Assumption 3 if $X$ causes $Y$) or the path contains $C$ or $S$ as a collider (which would imply $X$ causes $C$ or $S$, which is not the case given Assumptions 3 and 4 if $X$ causes $Y$).

We now start the proof of the Theorem with its second part (i.e., the claim that conditional d-separation of $Z_j$ and $Y$ implies that $C$ is a valid adjustment set).

First consider paths 1-4, where the path between $Z_j$ and $Y$ include $X$ (as a collider, which is the only possibility given our assumptions). Segment (a) denotes paths between $Z_j$ and $X$ that do not include $Y$, and segment (b) denotes backdoor paths between $X$ and $Y$ that do not include $Z_j$. In all cases, if the path between $X$ and $Y$ is open given $\{C_j^* \cup X\}$, then $Z_j$ and $Y$ will be conditionally d-connected. This is because the path between $Z_j$ and $X$ is open (by assumption) and $X$ is a collider. So, for $Z_j$ and $Y$ to be conditionally d-separated, conditioning on $C_j^*$ must be sufficient to block the paths between $Z_j$ and $Y$ – i.e., to block all backdoor paths between $X$ and $Y$ that do not include $Z_j$.

Now consider paths 5-12, which denote backdoor paths between $X$ and $Y$ that include $Z_j$. Segment (c) denotes paths between $Z_j$ and $Y$ that do not include $X$. First, consider all paths where $Z_j$ is not a collider (paths 5, 6, 7, 9, 11 and 12). If segment (c) is open given $C_j^*$ (although the conditioning set also includes $X$, such inclusion is not relevant because the path does not contain $X$), then $Z_j$ and $Y$ will be conditionally d-connected. So, for $Z_j$ and $Y$ to be conditionally d-separated, conditioning on $C_j^*$ must be sufficient to block segments (c) – i.e., to block all backdoor paths between $X$ and $Y$ that include $Z_j$.

Paths 8 and 10 denote backdoor paths between $X$ and $Y$ that include $Z_j$ as a collider and are therefore blocked at $Z_j$. In this case, even though $C_j^*$ is a valid adjustment set (assuming it blocks all other backdoor paths considered above), $Z_j$ and $Y_j$ are d-connected given $C_j^*$. Although this may at first seem misleading, this is actually the correct result, because the candidate adjustment set is $C$, not $C_j^*$. Since $Z_j \in C$, $C$ is not a valid adjustment set (because conditioning on $C$ requires conditioning on $Z_j$, which would open the backdoor path at $Z_j$) unless conditioning on $C_j^*$ also blocks segment (c) in paths 8 and 10.

Finally, we consider the case where the implication of Lemma 1 is satisfied by a path between $Z_j$ and $X$ that include $Y$ (paths 13-20). For $Z_j$ and $X$ to be d-connected conditional on $C_j^*$ through such paths, both segments (b) and (c) must be conditionally open. For paths 13 and 15, this also requires that $Y$ affects either $C_j^*$ or $S$, which is possible given our assumptions if $X$ does not cause $Y$. So, for $Z_j$ and $Y$ to be conditionally d-separated, conditioning on $C_j^*$ must be sufficient to block segment (c). Notice that, if only segment (c), but not segment (b), is blocked, there will be a backdoor path between $X$ and $Y$. But this would not lead to a wrong conclusion, because in this case $Z_j$ and $X$ will also be conditionally d-separated, and the Theorem's requirements would indicate that such $Z_j$ cannot be used for the proposed test.

Therefore, for $Z_j$ and $Y$ to be d-separated conditional on $\{C_j^* \cup X\}$, conditioning on $C_j^*$ must be sufficient to block all backdoor paths between $X$ and $Y$, whether or not such paths include $Z_j$. In this case, $C$ (and, indeed, $C_j^*$) would be a valid adjustment set. Of note, since this argument applies to any $Z_j$, if at least one $Z_j$ satisfies the above conditions, $C$ would be a valid adjustment set. This proves the second part of the Theorem.

The first part of the Theorem, with regards to $C$ being an invalid adjustment set, follows from the arguments above. If $C_j^*$ does not block segment (b), then $Z_j$ and $Y$ are conditionally d-connected (paths 1-4) and $C$ is not a valid adjustment set because such backdoor paths would also not be blocked by additional conditioning on $Z_j$. However, it is possible that conditioning on $C_j^*$ blocks segment (b) in paths 5, 6, 7, 9, 11 and 12, but not in paths 8 and 10. If paths 8 and 10 do not exist, then $C$ is a valid adjustment set, because conditioning on $Z_j$ would block the other paths. In this case, $Z_j$ and $Y$ would be conditionally d-connected even though $C$ is a valid adjustment set. In this situation, $C$ is minimally sufficient in the sense that neither $Z_j$ nor $C_j^*$ individually are valid adjustment sets. One example of this is when $C$ is a valid adjustment, but no proper subset of $C$ is a valid adjustment set. Regarding paths 12-20, if $C_j^*$ does not block segment (b), then $Z_j$ and $Y$ will be d-connected (otherwise, $Z_j$ and $X$ would be d-separated and such $Z_j$ would not satisfy the requirements of the Theorem). This proves the first part of the Theorem, thus concluding the proof.

As mentioned above, one limitation of the proposed approach is that, if $Z_j$ and $Y$ are dependent conditional on $\{C_j^* \cup X\}$ for all $j \in \{1, \dots, J\}$, it is unclear whether $C$ is a valid adjustment set or not. However, if $Z_j$ is known to be exogenous, then conditional d-connected between $Z_j$ and $Y$ implies $C_j^*$ is an invalid adjustment set. This is because, if $Z_j$ is exogenous, then paths 8 and 10 do not exist, thus eliminating the possibility that conditioning on $Z_j$ introduces bias.

## 2. Unconfoundedness assessment of time-varying exposures and in mediation analysis

Before discussing how to apply the method for exposure protocols involving multiple time points, we consider the meaning of Assumption 1 in the case where only a single measure of a time-varying exposure is of interest (as in the main text), both $X$ and $Y$ are time-varying and one may cause the other. In this case, measurement timing must be carefully considered. For example, let $X_t$ and $Y_{t'}$ represent the exposure and outcome variables measured at times $t$ and $t'$. If $t < t'$, then $Y_{t'}$ cannot be a cause of $X_t$, even though the set $\{Y_1, Y_2, \dots, Y_t\}$ may contain causes of $X_t$. In this case, such members of this set may be confounders of the effect of $X_t$ on $Y_{t'}$. This is different to the situation where $t = t'$, where both $X_t \to Y_{t'}$ and $Y_{t'} \to X_t$ are possible.

We now discuss how to apply the method to assess the plausibility of unconfoundedness when estimating the effect of exposure protocols over time. Let $t \in \{0,1, \dots, T\}$ index the time, and $Y = Y_{T+1}$ (to ensure Assumption 1 holds). So, for each $t$, we have the exposure $X_t$ and the candidate adjustment set $A_t$ (of note, $A_t$ may contain $X_{t'}$ and elements from $A_{t'}$, for all $t' < t$). Then, if assumptions 1-5 hold for all $t$, the test can be applied for each $X_t$ to assess conditional exchangeability. If the latter hold for all $X_t$, then sequential exchangeability holds and it is therefore possible to estimate the causal effect of exposure protocols involving multiple time points.[1] Similarly, our results can be applied to mediation analysis. Suppose we have an exposure measured at time $t = 0$ and a mediator $M$ measured at time $t = 1$. Then, $A_0$ denotes the candidate set for adjusting exposure-mediator and exposure-outcome confounding (also called base confounders); and $A_1$ denotes the candidate adjustment set for mediator-outcome confounding (also called post confounders).[2]

## 3. Assessing unconfoundedness in the presence of measurement error

We now explicitly discuss the influence of measurement error in the proposed method. For this, let $V'$ denote the measured version of some variable $V$. Now, let $W$ denote the (possibly theoretical) perfect version of the adjustment set, $T$ the (possibly theoretical) perfect version of the exposure and $G$ the (possibly theoretical) perfect version of the outcome. If covariates are measured with error, then $C = W' \neq W$. If the exposure is measured with error, then $X = T' \neq T$. If the outcome is measured with error, then then $Y = G' \neq G$. That is, $C$, $X$ and $Y$ denote the measured (possibly with error) versions of covariates.

An important consideration on measurement error is that the method and its assumptions are related to the measured versions of the variables, not to their theoretical perfectly measured counterparts. That is, the method aims at assessing whether the set of measured covariates is a valid adjustment set to estimate the effect of the measured version of the exposure on the measured version of the outcome. So, from this perspective, whether $C$, $X$ and/or $Y$ are mismeasured is irrelevant. Nevertheless, even though the question of whether $X = T'$ affects $Y = G'$ is well-defined, the question of scientific interest often relates to $T$ and $G$. From this perspective, measurement error is relevant because confounding between $X$ and $Y$ does not necessarily correspond to confounding between $T$ and $G$, and the association between $X$ and $Y$ is not necessarily a good approximation for the effect of $T$ on $G$ (as we discuss below). Therefore, the main implication of measurement error is not a validity issue (since measurement error does not influence the primary goal of our method, which relates to assessing confounding between $X$ and $Y$), but an interpretation issue (whether confounding between $X$ and $Y$ is indicative of confounding between $T$ and $G$).

In this section, we will consider two forms of measurement error: (i) classical error, which can be represented in DAGs as follows: $V \to V' \leftarrow V_C$, where $V_C$ represents sources of error in $V$;[1] and (ii) Berkson's error, which can be represented as $V' \to V \leftarrow V_C$.[3] Classical error happens when the true version of the variable is a cause of the measured version: for example, actual intake of a given medication is likely a cause of self-reported intake. Berkson's error happens when the true version of the variable is a consequence of the measured version: for example, $V$ could be a multicomponent construct (e.g., physical activity) and $V'$ a component of it (e.g., number of days the individual exercises in a typical week).

First, consider measurement error in covariates. In Supplementary Figures 2a (classical error) and 2b (Berkson's error), $Z$ and $Y$ would be d-connected given $X$ and $C = \{Z \cup D\}$ even though $W = \{Z \cup M\}$ is a valid adjustment set. This is the correct result because the actual adjustment set is $C \neq W$. Therefore, the proposed approach would correctly classify $C$ as an invalid adjustment set. More generally, it should be noted that Assumptions 1-4 refer to $C$ itself, so whether all elements in $C$ are perfectly measured versions of some set of covariates is irrelevant. This is a desirable property, since in practice interest lies in knowing whether the set of measured covariates is a valid adjustment set.

Now, consider measurement error in the outcome. Supplementary Figure 1c (classical error) and 2d (Berkson's error) each cover two scenarios: an unmeasured cause of $X$ (here, assumed to have no error), labelled $U$, also causes $G$ (indicated by the dashed arrow); and $U$ also causes $Y$ (indicated by the gray arrow). When considering the dashed arrow and ignoring the gray arrow, there is unmeasured confounding both between true exposure and true outcome and between measured exposure (in this case, equal to the true exposure) and measured outcome. In this situation, $Z$ and $Y$ are d-connected given $X$ (note $C = Z$). That is, in this situation, confounding between measured exposure and measured outcome corresponds to confounding between true exposure and true outcome, and this bias would be detectable by the proposed method. When considering the gray arrow and ignoring the dashed arrow), there is no unmeasured confounding between true exposure and true outcome, but there is unmeasured confounding between measured exposure and measured outcome. Again, $Z$ and $Y$ are d-connected given $X$. Therefore, in this situation, confounding between measured exposure and measured outcome does not correspond to confounding between true exposure and true outcome. Nevertheless, the proposed method would indicate bias, as expected, since (as explained above) its primary goal is to assess unmeasured confounding between measured exposure and measured outcome. Moreover, even though there is no confounding between true exposure and true outcome in this example, the association between measured exposure and measured outcome would be biased for the effect of the true exposure on the true outcome (note that $X$ and $Y$ are d-connected even though $X$ has no effect on $G$), so the fact that the proposed method would detect bias in this situation would be desirable.

When considering the dashed arrow and ignoring the gray arrow in Supplementary Figure 1d, there is unmeasured confounding between true exposure and true outcome, but no unmeasured confounding between measured exposure and measured outcome. $Z$ and $Y$ are d-separated given $X$. Therefore, in this situation, confounding between measured exposure and measured outcome does not correspond to confounding between true exposure and true outcome. In this case, the proposed method would not indicate bias, as expected, since its primary goal is to assess unmeasured confounding between measured exposure and measured outcome, which does not exist in this situation. Moreover, since the association between measured exposure and measured outcome would not be a good approximation for the effect of the true exposure on the true outcome (note that $X$ and $Y$ are d-separated even though $X$ causes $G$), the fact that the proposed method would not detect bias in this example would not be desirable. When considering the gray arrow and ignoring the dashed arrow, there is unmeasured confounding both between true exposure and true outcome and between measured exposure and measured outcome, and this would be detectable by the proposed method.

Now, we consider measurement error in the exposure. Supplementary Figures 2e (classical error) and 2f (Berkson's error) each cover two scenarios: an unmeasured cause of $Y$ (here, assumed to have no error), labelled $U$, also causes $T$ (indicated by the dashed arrow); and $U$ also causes $X$ (indicated by the gray arrow). We noticed the same patterns as those observed for classical and Berkson's error in the outcome: for example, when considering the dashed line and ignoring the gray line in Supplementary Figure 1e, there is unmeasured confounding both between measured and true versions of exposure and outcome, and the proposed method would detect this bias. This is the same pattern observed when considering the dashed line and ignoring the gray line in Supplementary Figure 1c (which is analogous to Supplementary Figure 1e since both consider classical error).

Now, we consider classical measurement error in both exposure and outcome. In this case, as illustrated in Supplementary Figure 2g, $Z$ and $Y$ would be d-connected given $X$ (note $C = Z$) if there is unmeasured confounding between $X$ and $Y$ (which can happen if there is unmeasured confounding between $T$ and $G$, as illustrated by the dashed arrows; or if there is unmeasured confounding between $X$ and $Y$ directly, as illustrated by the gray arrow). The fact that the proposed test would detect such confounding is desirable. However, $Z$ and $Y$ would be also d-connected given $X$ in Supplementary Figure 1h, but in this case this is happening due to unmeasured confounding between measured exposure and measured outcome due to a causal effect of true exposure on true outcome. If $T$ did not cause $G$ (i.e, if the dashed line were not in the DAG), then $Z$ and $Y$ would be d-separated given $X$ and $C$, as would be expected since there is no confounding between true exposure and true outcome. Since the primary goal of the proposed method is to detect unmeasured confounding between measured exposure and measure outcome, this is the correct result, because $T \notin C$ is a confounder of $X$ and $Y$ if $T$ causes $G$. However, such confounding by $T$ is what allows using the association between $X$ and $Y$ as a proxy for the association between $T$ and $G$. Therefore, from a practical perspective, it might have been desirable that the proposed approach did not detect this specific form of confounding.

The scenarios covered in Supplementary Figure 1, while not exhaustive, are sufficient to show that, although the proposed test allows (under Assumptions 1-5) assessing confounding in the association between measured exposure and measured outcome, extrapolating results to true exposure and true outcome requires assumptions about measurement error in exposure and outcome.

## 4. Comparison to previous methods

Pearl[4] describes several tests for confounding detection and explains their limitations. We now compare such tests with our proposed test to highlight advantages and disadvantages of each.

In his Definition 6.2.2, Pearl describes the following criterion: $X$ and $Y$ are classified as unconfounded if each member $M$ of $V$ (where $V$ denotes the set of variables in the problem that are not affected by $X$) satisfies at least one of the following conditions: $M$ is not associated with $X$ (condition $C_1$); $M$ is not associated with $Y$ conditional on $X$ (condition $C_2$). Otherwise (i.e., if any $M$ in $V$ violates both conditions), $X$ and $Y$ are classified as confounded. Of note, $M$, $V$, $C_1$ and $C_2$ respectively correspond to $Z$, $T$, $U_1$ and $U_2$ in Pearl's notation.

Using Pearl's terminology, a criterion for no-confounding is necessary if it is always correct when classifying $X$ and $Y$ as confounded, and sufficient if it is always correct when classifying $X$ and $Y$ as unconfounded. Pearl correctly describes that the criterion described above fails with respect to both properties. We now briefly describe how such failures can happen (an in-depth discussion is available elsewhere).[4]

Failure with respect to sufficient can happen via:

- Marginality: the associational criterion above tests each element of $V$ individually, but it could be the case that two factors jointly confound $X$ and $Y$ while each factor separately satisfies the condition for no confounding (i.e., each factors satisfies either $C_1$ or $C_2$). In this case, the criterion would incorrectly classify $X$ and $Y$ as unconfounded.
- Closed-world assumptions: the associational criterion classifies $X$ and $Y$ as unconfounded if all $M$ in $V$ satisfy either $C_1$ or $C_2$. However, it is possible that $V$ does not contain some confounders. So, unless $V$ contains all possible confounders (i.e., the closed-world assumption holds), $X$ and $Y$ can be confounded even though no $M$ in $V$ violate both conditions. Since this assumption is impossible to verify, Pearl argues that any confounding test will be insufficient.

Failure with respect to necessity can happen via:

- Barren proxies: $M$ can be a collider on a backdoor path between $X$ and $Y$, but nevertheless be associated with both $X$ and $Y$ (as $Z_j$ in path 8, or example). For example, consider a DAG which has a single path: $X \leftarrow U_1 \rightarrow M \leftarrow U_2 \rightarrow Y$ ($U_1$ and $U_2$ are unmeasured). In this case, $M$ violates both $C_1$ and $C_2$, but $X$ and $Y$ are not confounded.
- Incidental cancellations: it is possible, for example, that $M$ and $U$ (unmeasured) are causes of $X$ and $Y$, but their effects end up cancelling so that there is no net confounding. So, $M$ would violate both $C_1$ and $C_2$ even though $X$ and $Y$ are not confounded.

Pearl also describes a modified associational criterion (Definition 6.3.2) which excludes barren proxies from $V$ by assumption. However, the remaining limitations still apply.

Finally, Pearl proposes an operational test for stable no-confounding (Theorem 6.4.4). By stable no-confounding, it is meant that $X$ and $Y$ are structurally unconfounded (that is, $X$ and $Y$ are unconfounded regardless of the parameterization of the topological structure of the true causal model underlying the data). Of note, this is the same form of confounding our approach aims at detecting. Pearl's test is the following: let $M$ be a variable that is unaffected by $X$ but may possibly affect $Y$. If $M$ violates both $C_1$ and $C_2$, then $X$ and $Y$ are not stable unconfounded.

Pearl's test solves the failures with respect to necessity by focusing on stable confounding and by assuming that $M$ is not a barren proxy. Given the impossibility of ensuring the closed-world assumption holds (and therefore of sufficiency), Pearl's test focuses on detecting confoundedness – that is, if $M$ satisfies either $C_1$ or $C_2$, it is not necessarily the case that $X$ and $Y$ are stable unconfounded. Therefore, Pearl's test is not affected by insufficiency in the sense that it does not attempt to classify a case as unconfounded.

The last sentence above is the first important limitation of Pearl's test. That is, while it can be very useful for rejecting a given candidate adjustment set as sufficient for confounding adjustment, it can never provide positive evidence supporting that the candidate adjustment set is valid. In practice, researchers are of course interested in knowing whether their candidate adjustment set is wrong, but the key interest is in finding a valid candidate adjustment set, which would allow valid causal inferences.

We now describe how our proposed approach deals with the limitations of previous associational criteria and Pearl's operational test. Lack of sufficiency via marginality is not possible under faithfulness. To illustrate this, we use Pearl's example involving two members of $V$: $M_1 \sim \text{Benoulli}(0.5)$ and $M_2 \sim \text{Benoulli}(0.5)$, and $M_1$ and $M_2$ are independent. Then, $X = \mathbb{1}(M_1 = M_2)$ (where $\mathbb{1}$ denotes the indicator function) and $Y = \mathbb{1}(M_1 \neq M_2)$. In this example, $X$ and $Y$ are perfectly negatively correlated due to confounding, but $M_1$ and $M_2$ are not individually associated with $X$ or $Y$. One can note that this example is a specific parametrization of the model $X \sim \text{Benoulli}(p_X)$ and $Y \sim \text{Benoulli}(p_Y)$, where $p_X = \beta_0 + \beta_1 M_1 + \beta_2 M_2 + \beta_3 M_1 M_2$ (where $(\beta_0, \beta_1, \beta_2, \beta_3) \in \mathbb{R}^4\colon p_X \in [0,1]$) and $p_Y = \gamma_0 + \gamma_1 M_1 + \gamma_2 M_2 + \gamma_3 M_1 M_2$ (where $(\gamma_0, \gamma_1, \gamma_2, \gamma_3) \in \mathbb{R}^4\colon p_Y \in [0,1]$). In the example, $\beta_0 = \gamma_1 = \gamma_2 = 1$, $\beta_1 = \beta_2 = -1$, $\beta_3 = 2$, $\gamma_0 = 0$ and $\gamma_3 = -2$. Under this specific choice of parameters values, indeed there is no association of $M_1$ or $M_2$ with $X$ or $Y$. However, the overwhelming majority of possible values for these parameters would not result in lack of association.

The second possible mechanism for sufficiency violation is the closed-world assumption, which can never be guaranteed to hold. Our approach is not subject to this limitation because we only consider for the test members of $A$ that are conditionally d-connected with $X$ through a path directed into $X$. This, in combination with faithfulness, implies that any conditionally open path between $X$ and $Y$ directed into $X$ will result in an open path between the member of $A$ and $Y$ (see section 1 above for details). Therefore, one does not need to know in advance whether $A$ contains all possible confounders.

Regarding necessity, faithfulness implies that incidental cancellations are not possible, but this is not particularly relevant to us since, as Pearl, we are interested in stable unbiasedness. Moreover, our approach does not require assuming that $Z_j$ is not a barren proxy. This is because we are assessing whether $A$ (which includes $Z_j$) is a valid adjustment set (see section 1 above for details). To understand the importance of this feature of our test, consider Pearl's interpretation of the requirement that $M$ may possibly affect $Y$: in Pearl's view, such requirement implies the graph should not assume that such relationship does not exist – therefore, a directed path from $M$ to $Y$ must be included. Although there is nothing wrong with this interpretation, it is not a neutral one: it does imply that, under one or more combinations of values of parents of $Y$ (which may or may not exist in the study population), changing the value of $M$ through an intervention changes the value of $Y$. This reveals the difficulty of knowing in practice whether $M$ may possibly affect $Y$ in Pearl's sense: to not be willing to assume that $X$ does not cause $Y$ (in the sense that one does not know whether such causal effect exists or not) is different than assuming that $M$ causes $Y$ under some combination of values of parents of $Y$.

In practice, we believe that the most common situation is that researchers are not willing to make assumptions on either side. So, we employ the following weaker interpretation of "$M$ may possibly affect $Y$", which we believe to be more natural than Pearl's: while acknowledging that $M$ causing $Y$ as a possibility, one is also not willing to assume that $M$ causes $Y$ under some conditions – in other words, the researchers wants to refrain from making causal assumptions regarding the relationship between $M$ and $Y$. In this case, we should consider all scenarios compatible with a situation where we do not know whether $M$ affects $Y$ (i.e., scenarios with and without and arrow from $M$ to $Y$). Under this interpretation, $M$ is not guaranteed to not be a barren proxy, in which case Pearl's method may fail (because one of its requirements is violated) with respect to necessity.

So, in essence, ours and Pearl's tests have different assumptions: our test assumes faithfulness and make no assumptions on barren proxies; while Pearl's test assumes $M$ is not a barren proxy and does not require faithfulness. As a result, our test is sufficient but not necessary; and Pearl's test is necessary but not sufficient. However, a key difference is that, if we assume $Z_j$ is not a barren proxy, then our test is necessary when classifying $A_j^*$ as an invalid adjustment set (like Pearl's test, which makes no statement on whether $M$ is a valid adjustment set). However, if we supplement Pearl's test with faithfulness, the test does not become sufficient, because $M$ not violating $C_1$ or $C_2$ would not necessarily mean $X$ and $Y$ is unconfounded: for example, in a DAG having only the paths $M \to Y$ and $X \leftarrow U \to Y$ (where $U$ is unmeasured), $M$ would not violate $C_1$ even though there is confounding. Of note, such lack of sufficiency even under faithfulness is not surprising since the test was not formulated to provide positive evidence in support of unconfoundedness.

Another approach with substantial overlap with ours is the covariate selection method described by Entner and colleagues[5]. They describe two rules and show that, if one of them is satisfied, then the set of covariates being considered is a valid adjustment set. As in our paper, they assume it is known that $X$ does not cause any element of $V$, $Y$ is not a cause of $X$, and faithfulness. It is also assumed there is no selection whatsoever, which is much stronger than our assumption regarding selection (see Assumption 3 in section 2 in the main text).

To describe the rules, let $V$ be the set from which covariates for confounding adjustment are going to be selected. Moreover, let $W \in V$ (i.e., $w$ is one of the variables belonging to $V$), $K \subseteq V$ (i.e., $K$ is any subset of variables belonging to $V$) $D \subseteq V \cap W^C$ (i.e., $D$ is a subset of variables belonging to $V$ which does not include $W$). We can now define the two rules:

- Rule 1 states that, if there exists a $w$ and a $D$ such that (i) $W$ and $Y$ conditionally associated given $D$ and (ii) $W$ and $Y$ conditionally independent given $D$ and $X$, then $X$ causes $Y$ and $D$ is a valid adjustment set.
- Rule 2 comprises the following conditions: (i) $X$ and $Y$ are conditionally independent given $K$, (ii) $W$ and $X$ are conditionally dependent given $D$ and (iii) $W$ and $Y$ are conditionally independent $D$. If either (i) or both (ii) and (iii) are satisfied some $K$ or $W$- $D$ pair, then $X$ does not cause $Y$.

Under the assumptions outlined in section 2 in the main text, whenever our method classifies $A = \{Z, C\}$ as a valid adjustment set, Entner and colleagues' method will provide equivalent results. To see this, note that, under our assumptions, whenever $A$ is classified as a valid adjustment (which would also imply $C$ is a valid adjustment set, as discussed in section 2 in the supplement) set by our method, the following conditions hold: (a) there is at least one open path (conditional on $C$) between $Z$ and $X$ that points into $X$, and no conditionally open paths that do not point into $X$; (b) $Z$ and $Y$ are conditionally independent given $C$ and $X$ (this is the result that, according to the Theorem, implies that $A$ is a valid adjustment set). If $X$ causes $Y$, then (a) would imply that $Z$ and $Y$ are d-connected given $C$, because there is an open path between $Z$ and $Y$ through $X$. Under faithfulness, this would imply that condition (i) of rule 1 is satisfied. Since condition (ii) of rule 1 is the same as (b), and the latter is satisfied whenever our method classifies $A$ (and therefore $C$) as a valid adjustment set, then condition (ii) our rule 1 is trivially also satisfied in this situation. Therefore, if $X$ causes $Y$, whenever our method classified $A$ and $C$ as valid adjustment sets, rule 1 would be satisfied and $C$ would be classified as a valid adjustment set.

Now consider the case where $X$ does not cause $Y$. Note that condition (ii) of rule 2 is implied by (a) above under faithfulness. Indeed, the Theorem requires that condition (ii) of rule 2 is satisfied, with (a) being a consequence of it given the assumptions described in section 2 in the main text. If (b) is satisfied, our method will classify $A$ as a valid adjustment, and $X$ and $Y$ would be conditionally d-separated (and therefore independent) given $A$, indicating $X$ does not cause $Y$. Likewise, in this situation, then $Z$ and $Y$ will be conditionally d-separated given $C$ (because $C$ is a valid adjustment set, so any association between $Z$ and $Y$ must be mediated by $X$), thus satisfying condition (iii) of rule 2 and concluding that $X$ does not cause $Y$.

It should also be noted that Entner and colleagues' method may be able to classify a set of covariates as valid when ours cannot. For example, consider the simple scenario $X \leftarrow C \rightarrow Y$. In this case, $V = C$. Since $X$ does not cause $Y$, condition (i) of rule 2 would be satisfied. In contrast, since there $C$ would be associated with $Y$ given $X$ (there is no additional covariates in this example), which would be interpreted as an inconclusive result according to the Theorem. Similar examples in the case where $X$ does not cause $Y$ could be constructed to further illustrate this. The conclusion, therefore, is that our method can be seen as a special case of Entner and colleagues' method, since they arrive at equivalent conclusions when our method can classify $A$ as a valid adjustment set, while Entner and colleagues' method may be able to correctly identify a valid adjustment set even though our method would yield inconclusive results.

Despite the advantage of Entner and colleagues' method mentioned above, there are important strengths of our method. Our method only assesses one covariate set ($A$), while Entner and colleagues' method goes through all subsets of $A$, which would be computationally prohibitively even for relatively small sets of covariates. Although such exploration of subsets of $A$ could yield important results (for example, a subset of $A$ may be a valid adjustment set when $A$ as a whole is not), it would have at least two negative implications. One of them is the possibility of finding contradictory cases in finite samples, even though the two rules are mutually exclusive, which would lead to inconclusive results.[5,6] The second one is that going over all subsets may induce faithfulness violations. For example, consider a situation where the effect of $X$ on $Y$ is positive but some elements of $A$ are negative confounders (i.e., they induce a negative association between $X$ and $Y$) of different strengths (i.e., some are strong negative confounders, some are weaker negative confounders). Going through all subsets of $A$ in this case would essentially produce several negative confounding levels (depending on the covariates left out from the covariate set being tested), one of which may be close enough to the effect of $X$ on $Y$ so that condition (i) of rule 2 is satisfied in finite samples due to quasi-violations of faithfulness. In our implementation, the fact that we first select a subset of "testing" covariates (those associated with $X$ given all other covariates) facilitates assessing the plausibility of faithfulness (as illustrated and discussed in sections 5 and 6 in the main text). Finally, the selection of testing covariates does not involve $Y$ at all, and the association between $X$ and $Y$ plays no role in assessing if $A$ is a valid adjustment set. Conversely, Entner and colleagues' approach involve $Y$ when selecting the testing covariate (condition (i) of rule 1) and even the association between $X$ and $Y$ itself (which is the research question) is used for assessing the validity of the adjustment set (condition (i) of rule 2). Although this is not problematic in infinite samples satisfying the assumptions the method requires, in practice it may lead to preferential selection of covariates that yield the desired results – for example, if the hypothesis is that $X$ and $Y$ are causally unrelated, preference might be given to subsets of $A$ that satisfy condition (ii) in finite samples, which could be a result of faithfulness (quasi) violations, as discussed above. Therefore, our approach more clearly separates the analytical steps of covariate selection and causal effect estimation, which is generally considered a good practice for causal inference.[7]

Entner and colleagues'[5] also describe, in Theorem 4, a falsification test which essentially postulates that, if $Z$ is exogenous, associated with $X$ given $C$ and with $Y$ given $X$ and $C$, then $C$ is an invalid adjustment set. This is also the conclusion of our Theorem for the case of an exogenous $Z$ (see supplementary section 3 for details). However, as mentioned above, our method not only provide a falsification test (if $Z$ is exogenous) but can also correctly classify $A$ as a valid adjustment set.

**5. Simulation study**

As mentioned in the main text (section 4), we performed a series of simulations to corroborate the theoretical results. First, we provide a brief description of each scenario and the main results. Then, we provide a detailed description of each simulation scenario.

The R code used to perform simulations and other analyses is available at: https://github.com/FernandoHartwig/Confounding_test.

**5.1. Brief description and main results**

The first set of simulations involves two independent (of each other) measured covariates (i.e., $C = \{C_1, C_2\}$) and a single unmeasured confounder, where we assessed the power of the proposed approach to detect unmeasured confounding by testing the conditional association between $C_1$ and $Y$ given $X$ and $C_2$ using different P-value thresholds (for this simulation, we only consider $Z = C_1$). As shown in Supplementary Table 1, power was higher for larger sample sizes, stronger conditional associations between $C_1$ and $X$ and stronger unmeasured confounding.

The second set of simulations illustrates the importance of model misspecification. Again, there are two independent measured covariates. Again, we only consider $Z = C_1$. In scenario 2A, there is no unmeasured confounding, while in scenario 2B there is unmeasured confounding. In both scenarios, $C_1$ and $X$ are d-connected given $C_2$. Therefore, $C_1$ would be expected to satisfy condition (i) of the Theorem in both scenarios, but satisfy condition (ii) only in scenario A. Indeed, as shown in Supplementary Table 2, the expected results are observed when models for the conditional association between $C_1$ and $X$ (model 1) and for the conditional association between $C_1$ and $Y$ (model 2) are correctly specified. However, in scenario A, the rejection rate for a misspecified model 1 was slightly under 5% (which is the expected rejection rate under the null), implying that (apart from chance) the analysis fails to demonstrate that condition (i) holds. In this case, $C_1$ could not be used to assess unmeasured confounding, thus precluding identifying that $C$ is a valid adjustment set. In scenario B, a similar result was observed for a misspecified model 2, implying that $C_1$ would appear to satisfy both conditions (i) and (ii) of the Theorem, even though there is unmeasured confounding. Although these results are intentionally extreme due to designing the data-generating model so as to produce null results under model misspecification, they serve to illustrate the more general point that model misspecification may both reduce our capacity to detect that $C$ is a valid adjustment set (by reducing power to detect covariates that satisfy condition (i)) and to detect that $C$ is an invalid adjustment set (by reducing power to detect covariates that satisfy condition (i)). As with all regression analyses, careful checking for model misspecification is recommended.

The third set of simulations was more comprehensive and involved a larger number of covariates (20), varying the following aspects: correlation between covariates; different statistical cutoffs to classify whether a covariate satisfies conditions (i) and (ii); strength of confounding between $X$ and $Y$; true data-generating models for $X$ and $Y$; model specification; and number of covariates that are causes of only $X$ ($c_X$) or $Y$ ($c_Y$). Selected results are shown in Supplementary

Table 3. In scenario 3a, there is no bias in the association between $X$ and $Y$ under correct model specification. In this situation, if $c_X = 0$, then there is no covariate that satisfies both conditions (i) and (ii) even though unconfoundedness holds, and therefore the test has high rejection rates (as indicated in the L=0 columns, which is the proportion of simulated datasets where no covariates satisfy both conditions of the Theorem). This illustrates the ambiguity of the proposed test mentioned in section 3. However, if $c_X = 4$, then in this case there are four covariates that satisfy both conditions, and the rejection rate of the test is much lower. For incorrectly specified models, there is bias in the association between $X$ and $Y$, and again $c_X = 4$ has a smaller rejection rate compared to $c_X = 0$. This result is expected because $c_X = 4$ implies there are more covariates satisfying condition (i), and thus more opportunities to find at least one that also satisfies conditions (ii). Moreover, these four covariates have no direct effect on $Y$, so they are less likely to violate condition (ii) compared to confounders. Indeed, for $c_X = 4$, most simulated datasets had at least one covariate to satisfy both conditions (column L≥1), but this proportion decreases when considering datasets that had at least 2 or at least 3 covariates satisfying both conditions (columns L≥2 and L≥3). In scenario 3b, where less stringent statistical cutoffs were used, the rejection rate decreased both in the absence of bias (in which case, lower rejection rates are desired) and in the presence of bias (in which case, lower rejection rates are not desired). Indeed, in the results shown, the test performed very similarly between scenarios with and without bias in the association between $X$ and $Y$, unlike when a more stringent cutoff was used. Trends similar to those observed in scenario 3a regarding L were also observed when considering different data-generating models (scenario 3c) or when varying other aspects of the simulations. Nevertheless, the rejection rate varied notably depending on the underlying data-generating model, strength of confounding and correlation between covariates.

### 5.2. Detailed description

- Scenario 1:

Let $C_1$ and $C_2$ denote measured covariates, $U$ an unmeasured variable, $X$ the exposure and $Y$ the outcome. The data-generating model corresponding to scenario 1 was:

$U_i \sim N(0,1)$

${C_1}_i \sim N(0,1)$

${C_2}_i = \beta_{C_1 \to C_2} {C_1}_i + \varepsilon_{{C_2}_i}$, where $\varepsilon_{{C_2}_i} \sim N\left(0, \sigma^2_{C_2}\right)$

$X_i = \beta_{C_1 \to X} {C_1}_i + \beta_{C_2 \to X} {C_2}_i + \beta_U U_i + \varepsilon_{X_i}$, where $\varepsilon_{X_i} \sim N(0, \sigma^2_X)$

$Y_i = \beta_{C_2 \to X} {C_2}_i + \beta_U U_i + \varepsilon_{Y_i}$, where $\varepsilon_{Y_i} \sim N(0, \sigma^2_Y)$

$U_i$, ${C_1}_i$, $\varepsilon_{{C_2}_i}$, $\varepsilon_{X_i}$ and $\varepsilon_{Y_i}$ are mutually independent. The parameters $\sigma^2_{C_2}$, $\sigma^2_X$ and $\sigma^2_Y$ were chosen so that $\mathrm{Var}\left[{C_2}_i\right] = \mathrm{Var}[X_i] = \mathrm{Var}[Y_i] = 1$. In all simulations, $\beta_{C_1 \to C_2} = \beta_{C_2 \to X} = \beta_{C_2 \to Y} = 0.3$. The remaining parameters were sample size $n \in \{500{,}1000{,}5000{,}20000\}$, $\beta_{C_1 \to X} \in \{0.1{,}0.2{,}0.3{,}0.4{,}0.5\}$ and $\beta_U \in \{0{,}0.1{,}0.2{,}0.3{,}0.4{,}0.5\}$. The specific combinations of $n, \beta_{C_1 \to X}$ and $\beta_U$ we considered as shown in Supplementary Table 1.

In each of the 5000 simulated datasets (for each combination of parameters), we fitted the linear regression model $\hat{Y} = \gamma_0 + \gamma_1 C_1 + \gamma_2 C_2 + \gamma_3 X$ and obtained the P-value from a T test with $n - 4$ degrees of freedom for the null hypothesis that $\gamma_1 = 0$. Then, we estimated the rejection rate by calculating the proportion of simulated datasets where P<5%, P<1% and P<0.1%.

- Scenario 2:

Using the same notation as for scenario 1, the data-generating model corresponding to scenario 2 was:

$C_{1i} \sim N(0,1)$

$C_{2i} = 0.5U_i + \varepsilon_{C_{2}i}$, where $\varepsilon_{C_{2}i} \sim N\left(0, \sigma^2_{C_2}\right)$

In scenario 2A, the remaining variables were simulated as follows:

$$U_i = \text{sample}\left(\left\{-2 + \frac{4(j-1)}{n-1}\right\}_{j=1}^{n}\right)$$

$X_i = 0.5C_{1i} + 0.5U_i^2 + \varepsilon_{Xi}$, where $\varepsilon_{Xi} \sim N(0, \sigma^2_X)$

$Y_i \sim N(0,1)$

The $\text{sample}()$ operator means sampling without replacement from the sequence of numbers given as the argument, independently of all other variables in the model. So, $\{U_1, U_2, \dots, U_n\}$ were defined by shuffling a sequence of $n$ equally spaced numbers starting at -2 and ending at 2.

In scenario 2B, the remaining variables were simulated as follows:

$U_i \sim N(0,1)$

$X_i = 0.5C_{1i} + 0.5(U_i + 0.2)^2 + \varepsilon_{Xi}$, where $\varepsilon_{Xi} \sim N(0, \sigma^2_X)$

$Y_i = 0.3269946(U_i + 0.1417565)^2 + \varepsilon_{Yi}$, where $\varepsilon_{Yi} \sim N(0, \sigma^2_Y)$

$U_i$, $C_{1i}$, $\varepsilon_{C_{2}i}$, $\varepsilon_{Xi}$ and $Y_i$ (scenario 2A) or $\varepsilon_{Yi}$ (scenario 2B) are mutually independent. The parameters $\sigma^2_{C_2}$, $\sigma^2_X$ and $\sigma^2_Y$ were chosen so that $\text{Var}\left[C_{2i}\right] = \text{Var}[X_i] = \text{Var}[Y_i] = 1$. Sample size was set to $n$ =20,000.

In each of the 5000 simulated datasets (for each combination of parameters), we fitted four linear regression models:

Model 1, linear: $\hat{X} = \alpha_0 + \alpha_1 C_1 + \alpha_2 C_2$

Model 1, quadratic: $\hat{X} = \beta_0 + \beta_1 C_1 + \beta_2 C_2^2$

Model 2, linear: $\hat{Y} = \gamma_0 + \gamma_1 C_1 + \gamma_2 C_2 + \gamma_3 X$

Model 2, quadratic: $\hat{Y} = \delta_0 + \delta_1 C_1 + \delta_2 C_2^2 + \delta_3 X$

Note that the name given to each model above corresponds to the classification used in Supplementary Table 2.

From each model, we obtained the P-value from a T test with $n-3$ (model 1) or $n-4$ (model 2) degrees of freedom for the null hypothesis that the coefficient for $C_2$ ($\alpha_2$, $\beta_2$, $\gamma_2$ and $\delta_2$) is 0. We also obtained the regression coefficient for $X$ ($\gamma_3$ and $\delta_3$). Then, for each of these four statistical tests, we estimated the rejection rate by calculating the proportion of simulated datasets where P<5%. Finally, we estimated the bias of the regression coefficient of $X$ as an estimator of the causal effect of $X$ on $Y$ by averaging value of the coefficient for $X$ over the simulated datasets.

- Scenario 3:

Let $C = (C_1, \dots, C_{20})$ denote a vector of covariates (indexed by $j \in \{1, \dots, 20\}$), $X$ the exposure and $Y$ the outcome. The data-generating model corresponding to scenario 3 was:

$$M_i \sim N(\mu, \Omega)$$

$$C_{j_i} = \begin{cases} M_i & \text{if } j \in \{1, \dots, 10\} \\ \mathbb{1}(M_i \geq 0) & \text{if } j \in \{11, \dots, 20\} \end{cases}$$

$$X_i = \left(\sqrt{\theta_X}/\mathrm{sd}(X_i^*)\right)X_i^* + \left(\sqrt{\theta_X}/\left(12\mathrm{sd}(X_i')\right)\right)X_i' + \varepsilon_{X_i}, \text{ where } \varepsilon_{X_i} \sim N(0, \sigma_X^2)$$

$$Y_i = \left(\sqrt{\theta_Y}/\mathrm{sd}(Y_i^*)\right)Y_i^* + \left(\sqrt{\theta_Y}/\left(12\mathrm{sd}(Y_i')\right)\right)Y_i' + \varepsilon_{Y_i}, \text{ where } \varepsilon_{Y_i} \sim N(0, \sigma_Y^2)$$

In this simulation model, $\mu$ is a 20-dimensional vector of zeros and $\Omega$ is a matrix where all diagonal elements are 1 and all off-diagonal elements are $\rho$. $\mathbb{1}$ denotes the indicator function, which is used to dichotomize half of the covariates.

Let $\alpha_X = \left(\alpha_{X_1}, \dots, \alpha_{X_{10}}\right)$, $\beta_X = \left(\beta_{X_1}, \dots, \beta_{X_{10}}\right)$ and $\gamma_X = \left(\gamma_{X_1}, \dots, \gamma_{X_{10}}\right)$ be vectors, defined as follows:

$$\alpha_{X_j} \sim \mathrm{Unif}(0.01, 0.03)$$

$$\beta_{X_j} = 8\alpha_{X_j}$$

$$\gamma_{X_j} = -2\beta_{X_j}$$

Define other sets of vectors as follows:

$$\alpha_{X_i}^* = \alpha_{X_j} I_j^* \text{ for } j \in \{1, \dots, 10\}$$

$$\beta_{X_i}^* = \beta_{X_j} I_{j+10}^* \text{ for } j \in \{1, \dots, 10\}$$

$$\gamma_{X_i}^* = \gamma_{X_j} I_j^* \text{ for } j \in \{1, \dots, 10\},$$

where $I_j^* = 1$ indicates that covariate $j$ is a confounder of the association between $X$ and $Y$ (0 otherwise). $I_j^*$ was defined as follows:

$$I_j^* = \mathrm{sample}\left(\left\{\{1\}_{k=1}^{12}, \{0\}_{m=1}^{8}\right\}\right),$$

where the $\mathrm{sample}()$ operator is defined as described in scenario 2 above. Of the 20 covariates, 12 have a direct effect on both $X$ and $Y$ and are therefore confounders.

Define additional sets of vectors as follows:

$$\alpha_{X_i}' = \alpha_{X_j} I_{X_j}' \text{ for } j \in \{1, \dots, 10\}$$

$$\beta_{X_i}' = \beta_{X_j} I_{X_{j+10}}' \text{ for } j \in \{1, \dots, 10\}$$

$$\gamma_{X_i}' = \gamma_{X_j} I_{X_j}' \text{ for } j \in \{1, \dots, 10\},$$

where $I'_{X\,j} = 1$ indicates that covariate $j$ has a direct effect on $X$, but is not a confounder (0 otherwise). $I'_{X\,j}$ was defined as follows:

$$I'_{X\,j} = \begin{cases} 0 & \text{if } j \in \left\{f \in \{1, \dots, 20\} | I^*_f = 1\right\} \\ \text{sample}\left(\{1\}_{k=1}^{c_X}, \{0\}_{m=1}^{8-c_X}\right) & \text{otherwise} \end{cases}.$$

Of the 20 covariates, $c_X$ have a direct effect on $X$ but not on $Y$.

Now, $X^*_i$ and $X'_i$ can be explicitly described:

$$\tilde{X}^*_i = \sum_{j=1}^{10} \left(\alpha^*_{X\,i} {C_j}_i^3 + \beta^*_{X\,i} {C_{j+10}}_i + \gamma^*_{X\,i} {C_j}_i {C_{j+10}}_i\right)$$

$$X^*_i = f_X\left(\tilde{X}^*_i\right)$$

$$\tilde{X}'_i = \sum_{j=1}^{10} \left(\alpha'_{X\,i} {C_j}_i^3 + \beta'_{X\,i} {C_{j+10}}_i + \gamma'_{X\,i} {C_j}_i {C_{j+10}}_i\right)$$

$$X'_i = f_X\left(\tilde{X}'_i\right),$$

where $f_X$ is some function that varies between scenarios within scenario 3.

Similarly, let $\alpha_Y = \left(\alpha_{Y\,1}, \dots, \alpha_{Y\,10}\right)$, $\beta_Y = \left(\beta_{Y\,1}, \dots, \beta_{Y\,10}\right)$ and $\gamma_Y = \left(\gamma_{Y\,1}, \dots, \gamma_{Y\,10}\right)$ be vectors, defined as follows:

$$\alpha_{Y\,j} \sim \text{Unif}(0.01, 0.03)$$

$$\beta_{Y\,j} = 8\alpha_{Y\,j}$$

$$\gamma_{Y\,j} = -2\beta_{Y\,j}$$

Define other sets of vectors as follows:

$\alpha^*_{Y\,i} = \alpha_{Y\,j} I^*_j$ for $j \in \{1, \dots, 10\}$

$\beta^*_{Y\,i} = \beta_{Y\,j} I^*_{j+10}$ for $j \in \{1, \dots, 10\}$

$\gamma^*_{Y\,i} = \gamma_{Y\,j} I^*_j$ for $j \in \{1, \dots, 10\}$,

$\alpha'_{Y\,i} = \alpha_{Y\,j} I'_{Y\,j}$ for $j \in \{1, \dots, 10\}$

$\beta'_{Y\,i} = \beta_{Y\,j} I'_{Y\,j+10}$ for $j \in \{1, \dots, 10\}$

$\gamma'_{Y\,i} = \gamma_{Y\,j} I'_{Y\,j}$ for $j \in \{1, \dots, 10\}$,

where $I'_{Y\,j} = 1$ indicates that covariate $j$ has a direct effect on $Y$, but is not a confounder (0 otherwise). $I'_{Y\,j}$ was defined as follows:

$$I_{Y'\,j} = \begin{cases} 0 & \text{if } j \in \left\{f \in \{1, \dots, 20\} | I^*_f + I'_{X\,f} = 1\right\} \\ \text{sample}\left(\{1\}_{k=1}^{c_Y}, \{0\}_{m=1}^{8-c_X-c_Y}\right) & \text{otherwise} \end{cases}.$$

Of the 20 covariates, $c_Y$ have a direct effect on $Y$ but not on $X$.

Now, $Y_i^*$ and $Y_i'$ can be explicitly described:

$$\tilde{Y}_i^* = \sum_{j=1}^{10} \left( \alpha_{Y_i}^* {C_j}_i^3 + \beta_{Y_i}^* C_{j+10_i} + \gamma_{Y_i}^* C_{j_i} C_{j+10_i} \right)$$

$$Y_i^* = f_Y\left(\tilde{Y}_i^*\right)$$

$$\tilde{Y}_i' = \sum_{j=1}^{10} \left( \alpha_{Y_i}' {C_j}_i^3 + \beta_{Y_i}' C_{j+10_i} + \gamma_{Y_i}' C_{j_i} C_{j+10_i} \right)$$

$$Y_i' = f_Y\left(\tilde{Y}_i'\right),$$

where $f_Y$ is some function that varies between scenarios within scenario 3.

Note that $X_i$ and $Y_i$ consist of adding scaled versions of $X_i^*$ and $X_i'$, and $Y_i^*$ and $Y_i'$ ($\mathrm{sd}()$ denotes the sample standard deviation), respectively, plus an error term with variance $\sigma_X^2$ and $\sigma_Y^2$ (respectively) chosen such that $\mathrm{Var}[X_i] = \mathrm{Var}[Y_i] = 1$. This implies that the 20 covariates collectively explain $100\left(\frac{12+c_X}{12}\right)\theta_X\%$ (for $\theta_X \in [0,1]$) of the variance of $X_i$ in a correctly specified model and $100\left(\frac{12+c_Y}{12}\right)\theta_X\%$ (for $\theta_Y \in [0,1]$) of the variance of $Y_i$ in a correctly specified model. The larger the value of $\theta$, the stronger the confounding between $X$ and $Y$. $M_i$, $\varepsilon_{X_i}$, $\varepsilon_{Y_i}$, $\alpha_{X_j}$ and $\alpha_{Y_j}$ are mutually independent.

As shown in Supplementary Table 3, scenario 3 includes 7 sub-scenarios (3a-3f), totaling 21 unique combinations of parameters. For each combination, 5000 datasets, each containing 5000 observations, were simulated. We estimated the bias of the regression coefficient of $X$ as an estimator of the causal effect of $X$ on $Y$ by averaging value of the coefficient for $X$ over the simulated datasets. We also calculated the proportion of simulated datasets where no covariates were classified as satisfying both conditions (i) and (ii) of the Theorem; and the proportion of simulated datasets where at least 1, at least 2 and at least 3 covariates were classified as satisfying both conditions (i) and (ii).

We now provide additional explanations for selected simulation parameters:

(i) True data-generating models. We varied the true data-generating models through $f_X$ and $f_Y$. More specifically, $f_X$ could be one of the following: identity function (i.e., $f_X(V_i) = V_i$, where $V_i$ denotes a generic random variable), exponential function (i.e., $f_X(V_i) = e^{V_i}$) and logarithmic function (i.e., $f_X(V_i) = \ln(V_i - \min(V_i) + 2)$, where $\min()$ denotes the sample minimum). The same possibilities were used for $f_Y$.

(ii) Model specification. For implementing the confounding assessment method, we considered three sets of models. (1) Linear regression models for $X$ given $C$ and for $Y$ given $X$ and $C$ that are correctly specified if $f_X$ and $f_Y$ are identity functions. (2) Linear regression models for $X$ and $Y$ including all 20 covariates but not including product terms and continuous covariates were not cube-transformed. Therefore, bias due to model misspecification is expected even if $f_X$ and $f_Y$ are identity functions. (3) Similar to model speficication 2, except that 6 confounders (selected at random) are missing from the covariate set. Therefore, bias due to both model misspecification and unmeasured confounding is expected even if $f_X$ and $f_Y$ are identity functions.

(iii) Statistical cutoff stringency. We considered two sets of statistical criteria to assess conditions (i) and (ii) of the theorem. A less stringent set, where a given covariate was classified as satisfying condition (i) if P-value<0.05/20 (model specifications 1 and 2) or P-value<0.05/14 (model specification 3) from the linear regression model for X given C (see below for details on how P-values were computed) – i.e., a Bonferroni-corrected threshold of 5%; and as satisfying condition (ii) if P-value≥0.05 from the linear regression model for Y given X and C. Note that condition (ii) did not consider multiple testing correction, which is conservative in this case because satisfying condition (ii) would indicate no residual/unmeasured confounding. We also considered a more stringent set, where the P-value thresholds were P-value<0.001/20 or P-value<0.001/14 for condition (i), and P-value≥0.3 for condition (ii).

P-values for covariate $j$ were computed for the null hypothesis that all regression coefficients related to covariate $j$ are zero. For model specifications 2 and 3, there is only one such coefficient ($h = 1$). For models 1, there are two such coefficients ($h = 2$): the coefficient for the "main effect" term and the coefficient for the product term. Therefore, P-values were computed by an ANOVA comparison of two linear regression models: the full model, containing all coefficients ($d = 31$ for condition (i) and $d = 32$ for condition (ii) for model specification 1, $d = 21$ and $d = 22$ for specification 2, and $d = 15$ and $d = 16$ for model specification 3); and the reduced model, containing all but the coefficients related to covariate $j$ ($d = 29$ and $d = 30$ for model specification 1, $d = 20$ and $d = 21$ for specification 2, and $d = 14$ and $d = 15$ for model specification 3). More specifically, P-values were obtained from an F test with $(h, n - d)$ degrees of freedom.

## 6. Applied example: model selection

As mentioned in the main text, we explored different strategies for modeling (1) the association of exposure with covariates, and (2) the association of outcome with exposure and covariate. In this section we describe this process in detail.

For (1), we considered the following logistic regression models for $X$:

1.1: This includes all covariates with no higher order terms nor transformations.

1.2: Same as 1.1, plus product terms between all pairs of covariates and quadratic terms for numeric covariates.

1.3: This includes only a single categorical variable generated by conditional inference tree analyses using the Akaike information criterion (AIC) as a stopping criterion (details below).

1.4: Same as 2.4 but using the Bayesian information criterion (BIC) instead of the AIC.

1.5: Combines 1.1 and 1.3.

1.6: Combines 1.1 and 1.4.

For the conditional inference tree analyses, partitions were performed until the AIC or BIC started to increase, under the constraint that no terminal node could contain less than 1% of the observations. This was performed by varying the p-value threshold for partition as follows: 0.90, 0.89, …, 0.02, 0.01, 0.005, 0.0025, 0.00125, …. For each threshold, a categorical variable composed of the terminal nodes of the resulting tree was obtained, and the AIC or BIC associated with such variable was calculated from a logistic regression model of $X$ on such categorical variable.

For (2), we considered the following linear regression models for $Y$:

2.1: This includes all covariates and exposure with no higher order terms nor transformations.

2.2: Same as 2.1, plus product terms between all pairs of covariates (but not between exposure and covariates) and quadratic terms for numeric covariates.

2.3: Same as 2.1, plus product terms between the exposure and each covariate.

2.4: This includes a single categorical variable generated by conditional inference tree analyses using the AIC as a stopping criterion plus the exposure, with no product terms.

2.5: Same as 2.4 but using the BIC instead of the AIC.

2.6: Same as 2.3, plus a product term between the categorical variable and the exposure.

2.7: Same as 2.4, plus a product term between the categorical variable and the exposure.

Conditional inference trees were performed as described above, except that linear instead of logistic regression was used.

We used 50 iterations of 5-fold cross validation for model selection. The result of each iteration was the average of the log-likelihood of each model over the 5-fold cross validation procedure. The result was the average of the mean log-likelihoods over the 50 iterations. The cross-validation process was repeated 50 times to minimize the influence of how the data is partitioned. The model that yielded the smallest log-likelihood (averaged over iterations and cross-validation processes) was selected and assumed to be a good estimate of the true model.

## 7. Description and assessment of power calculations

This section is structured as follows. First, we describe how the confounding assessment method was implemented in the applied example. Second, we describe how the power of the confounding assessment method in our applied example was estimated. Third, we describe how the performance of the proposed power calculation approach in our applied example was assessed.

- Implementation of the confounding assessment method in the applied example:

Let $X$, $Y$ and $C$ denote the real exposure (delivery mode), outcome (intelligence quotient) and the candidate adjustment set comprised of eight covariates, respectively. As described in the main text (section 5), the confounding assessment method was implemented in the applied example by fitting a logistic regression model for $X$ given $C$ and a linear regression model for $Y$ given $X$ and $C$. None of these models included transformations or product terms, which was considered the best model by the model selection procedure described in detail in supplementary section 7. These models were used to obtain P-values for the conditional association between each covariate with $X$ and $Y$, which were then used to classify each covariate as satisfying (or not) conditions (i) and (ii) of the Theorem, respectively.

More specifically, P-values for the conditional association between covariate $j$ and $X$ were obtained using a likelihood ratio test implemented as a Chi-squared test, comparing a model including all covariates (“full model”) with a model excluding covariate $j$ (“reduced model”). Covariate j was classified as satisfying condition (i) of the Theorem if this P-value was less than 0.05.

P-values for the conditional association between covariate $j$ and $Y$ were obtained similarly, except that full and reduced models were compared using an ANOVA test, with P-values calculated using a F-test. Covariate j was classified as satisfying condition (ii) of the Theorem if this P-value was at least 0.3.

- Power calculations:

We now describe in detail the power calculation approach (using plasmode simulations) used to estimate the power of the confounding assessment method to detect unmeasured confounding in the real dataset analyzed in section 5 in the main text.

Let $\tilde{X}$ and $\tilde{Y}$ denote simulated exposure and outcome, respectively. Moreover, $\tilde{U}$ denotes a simulated unmeasured variable.

To incorporate sampling uncertainty associated with the real dataset, the plasmoid simulations were performed by non-parametrically bootstrapping the real dataset and simulating variables for each bootstrap iteration, each indexed by $r \in \{1, \dots, R\}$, where $R$ denotes the number of bootstrap iterations.

Within each bootstrap iteration, we simulate as follows:

$\tilde{U}_{r_i} \sim Bernoulli(p_U)$, where $p_U = 0.5$

$\tilde{X}_{r_i} \sim Bernoulli(\pi_{r_i})$, where $\pi_{r_i} = \text{sigmoid}(\gamma_r + \rho_{r_i} + \alpha_X \tilde{U}_{r_i})$

$\tilde{Y}_{r_i} = \alpha_Y \tilde{U}_{r_i} + {\beta_Y}^T C_{r_i} + \varepsilon_{r_i}$, where $\varepsilon_{r_i} \sim N(0, \sigma_r^2)$

Note that the subscript $r$ indicates the version of the variable in iteration $r$. So, for example, $\tilde{X}_{r_i}$ denotes the $i$-th observation of the simulated exposure in the bootstrapped dataset in iteration $r$. $C_{r_i}$ denotes the $i$-th observation of the real covariate vector in the bootstrapped dataset in iteration $r$.

As mentioned above, neither the logistic model for $X$ given $C$ nor the linear regression model for $Y$ given $X$ and $C$ included transformations or product terms. Therefore, the models for $\tilde{X}_r$ and $\tilde{Y}_r$ also did not include such terms, as we now describe.

For $\tilde{X}_r$, $\rho_r$ denotes the predicted values (on the ln-odds scale) of a logistic regression model for $X_r$ given $C_r$, specified as the model for $X$ given $C$. Including $\rho_r$ allows incorporating the observed pattern of statistical associations between covariates and exposure in the simulated exposure (using $X_r$ and $C_r$ instead of $X$ and $C$ to incorporate sampling uncertainty), which is also influenced by an unmeasured factor and a constant term $\gamma_r$, chosen so that the sample proportion of $\tilde{X}_r = 1$ is (almost) the same as the sample proportion of $X_r = 1$. The direct effect of $\tilde{U}_r$ on $\tilde{X}_r$, in the ln(odds ratio) scale, is given by $\alpha_X = \ln\left(\frac{\tau_1/(1-\tau_1)}{\tau_0/(1-\tau_0)}\right)$, where $\tau_1 = p_U + (1 - p_X)\frac{\delta}{\alpha_Y}$, $p_X$ is the sample proportion of $X = 1$, and $\delta$ is the (expected) bias of the coefficient for $\tilde{X}_r$ in a linear regression model of $\tilde{Y}_r$ given $\tilde{X}_r$ and $C_r$ as an estimator of the causal effect of $\tilde{X}_r$ on $\tilde{Y}_r$, which is zero; and $\tau_0 = \tau_1 - \frac{\delta}{\alpha_Y}$.

For $\tilde{Y}_r$, $\sigma_r^2$ was chosen so that $\text{Var}[\tilde{Y}_r] = 1$, which is the sample variance of $Y$ since the outcome was standardized. The direct effect of $\tilde{U}_r$ on $\tilde{Y}_r$, in the mean difference scale, is given by $\alpha_Y = \sqrt{\frac{\varphi}{p_X(1-p_X)}}$, where $\varphi$ is the proportion of the variance of $\tilde{Y}_r$ explained by the direct effect of $\tilde{U}_r$ on $\tilde{Y}_r$. For incorporating the association between covariates and outcome in the simulated outcome (i.e., defining $\beta_Y$), we used two approaches. The first used the coefficients from a linear regression model of $Y$ given $X$ and $C$; the second used coefficients from a linear regression model of $Y$ given $C$ only. As justified above, none of these models included transformations or product terms. The first strategy is prone to bias if there is unmeasured confounding in the real dataset, in which case $X$ is a collider between covariates satisfying condition (i) of the Theorem and the unmeasured confounder; the second is prone to bias if the effect of $X$ on $Y$ is not zero, in which case the coefficients for $C$ would include both the direct association between the covariate and the outcome and the indirect association mediated by the effect of $X$. Since the two approaches have limitations, including both is important to assess if the results of the power calculations are sensitive to this choice.

The models from each approach were filtered by backward selection, where the covariate with the largest P-value was removed until doing so reduced the Bayesian Information Criterion (BIC) of the model. This was performed to mitigate the risk of spuriously inflating the capacity of the method to detect unmeasured confounding and results in the method being conservative, as shown in the simulations designed to assess the performance of the power calculations (described below). After filtering, the coefficients of the covariates that remained in the model were assigned to the corresponding positions in $\beta_Y$; for others, the corresponding positions in $\beta_Y$ were set to zero.

The confounding assessment test was then applied to the simulated data. More specifically, among covariates that were classified as satisfying condition (i) of the Theorem in the real dataset, we checked which ones were also classified as satisfying condition (ii) in the simulated data. For this, a linear regression model for $\tilde{Y}_r$ given $\tilde{X}_r$ and $C_r$ (with no transformations or product terms) was fitted, and P-values obtained as described above. Of note, we used the real rather than the simulated dataset for condition (i) to retain as much information directly obtained from the real data as possible. Indeed, whether an unmeasured confounder that is independent of the covariates exists is not expected to have any influence on the association between covariates and the exposure, only on the association between covariates and outcome given the exposure.

A total of six scenarios were evaluated in the power calculations by taking all six possible combinations of $\delta \in \{0.2,0.4\}$ and $\varphi \in \{0.1,0.2,0.4\}$. For each combination, $R = 5000$. Since the confounding assessment method, when applied to the real dataset, classified only one covariate as satisfying conditions (i) and (ii) of the Theorem, we estimated the probability of classifying at least one covariate as satisfying these two conditions as the proportion of bootstrap iterations where at least one covariate is classified as satisfying them. If this proportion is high, then it is likely that at least one covariate is classified as satisfying both conditions of the Theorem for a given level of unmeasured confounding.

- Assessment of the performance of the power calculation method:

We now describe in detail the simulations used to assess the performance of the power calculation approach (described above) in the real dataset analyzed in section 5 in the main text.

The goal of this analysis is to (a) use simulated data to estimate the true performance of the power calculations in this particular data-generating model; and (b) compare these estimates to those obtained by applying the power calculation method described above to the simulated data. Step (a) was performed via plasmode simulations so that the data-generating model resembles the real dataset as much as possible. We now describe step (a) in detail.

Let $\overline{U}_t$, $\overline{X}_t$ and $\overline{Y}_t$ respectively denote the simulated unmeasured confounder, simulated exposure and simulated outcome in iteration $t \in \{1, \dots, T\}$, where $T$ is the number of simulated datasets for a given combination of simulation parameters.

Within each iteration, we simulate as follows:

$\overline{U}_{t_i} \sim Bernoulli(p_U)$, where $p_U = 0.5$

$\overline{X}_{t_i} \sim Bernoulli(\pi_{t_i})$, where $\pi_{t_i} = \mathrm{sigmoid}(\gamma_t + \rho_i + \alpha_X \overline{U}_{t_i})$

$\overline{Y}_{t_i} = \zeta \overline{X}_t + \alpha_Y \overline{U}_{t_i} + \omega_Y{}^T C_i + \varepsilon_{t_i}$, where $\varepsilon_{t_i} \sim N(0, \sigma_t^2)$

Note that the subscript $t$ indicates the version of the variable in iteration $t$. So, for example, $\overline{X}_{t_i}$ denotes the $i$-th observation of the simulated exposure in iteration $t$. $C_i$ denotes the $i$-th observation of the real covariate vector, since covariates were not simulated. Importantly, this should not be confused with the bootstrap procedure described above. In step (a), we are simulating the data that will the treated as the observed data for the bootstrap procedure in step (b).

For $\overline{X}_t$, $\rho$ denotes the predicted values (in the ln-odds scale) of a logistic regression model for $X$ given $C$. Including $\rho$ allows incorporating the observed pattern of statistical associations between covariates and exposure in the simulated exposure, which is also influenced by an unmeasured factor and a constant term $\gamma_t$, chosen so that the sample proportion of $\overline{X}_t = 1$ is (almost) the same as the sample proportion of $X = 1$. The direct effect of $\overline{U}_t$ on $\overline{X}_t$, in the ln(odds ratio) scale, is given by $\alpha_X = \ln\left(\frac{\tau_1/(1-\tau_1)}{\tau_0/(1-\tau_0)}\right)$, where $\tau_1 = p_U + (1 - p_X)\gamma$, $p_X$ is the sample proportion of $X = 1$ and $\tau_0 = \tau_1 - \gamma$. The parameter $\gamma$ captures the strength of the association between $\overline{U}_t$ and $\overline{X}_t$. In our simulations, $\gamma$ ranged from smaller to larger values, as specified below.

For $\overline{Y}_t$, $\sigma_t^2$ was chosen so that $\mathrm{Var}[\overline{Y}_t] = 1$, which is the sample variance of $Y$ since the outcome was standardized. The direct effect of $\widetilde{U}_r$ on $\widetilde{Y}_r$, in the mean difference scale, is given by $\alpha_Y = \frac{\delta}{\gamma}$, where $\delta$ is the (expected) bias of the coefficient for $\overline{X}_t$ in a linear regression model of $\overline{Y}_t$ given $\overline{X}_t$ and $C$ as an estimator of the causal effect of $\overline{X}_t$ on $\overline{Y}_t$, which is given by $\zeta$. Of note, $\alpha_X$ and $\alpha_Y$ are inversely related, since if the association between $\overline{U}_t$ and $\overline{X}_t$ is weak (i.e., $\gamma$ is small), then $\alpha_Y$ must be relatively big to ensure the desired bias $\delta$. For defining $\omega_Y$ (the direct effect vector of $C$ on $\overline{Y}_t$), we first estimated coefficients of a linear regression model for $Y$ given $X$ and $C$. Then, $8 - c_Y$ covariates were selected at random (so different sets of covariates were selected in different iterations), and all regression coefficients corresponding to these covariates were set to zero. This way, we could regulate the number of covariates that had a direct effect on $\overline{Y}_t$, and such effects are compatible with the conditional associational coefficients observed in the real data. This set of coefficients was treated as the true $\omega_Y$.

We then applied the confounding assessment method, as described above, to $\overline{X}_t$, $\overline{Y}_t$ and $C$. By performing this over $T = 1000$ iterations, we estimated the true performance of the confounding assessment method by calculating the proportion of simulated datasets where at least 1, 2 and 3 covariates were classified as satisfying both condition (i) and condition (ii) of the Theorem.

We now describe step (b). Essentially, in each iteration $t$, we used $\bar{X}_t$ and $\bar{Y}_t$ instead of $X$ and $Y$, respectively, for the power calculations. That is, we applied the power calculation approach described above, but using simulated instead of real data. Given the high computational burden, for this analysis we set $R = 200$ instead of $R = 5000$. Importantly, we used the notation $\alpha_X$ to denote both the direct effect of $\tilde{U}_r$ on $\tilde{X}_r$ and the direct effect of $\bar{U}_t$ on $\bar{X}_t$. This emphasizes that the direct effect of the unmeasured confounder on the exposure assumed is assumed to be correct – that is, the direct effect assumed in the power calculations is the same as the direct effect in the data-generating model. Similarly, $\alpha_Y$ was used to denote both the direct effect of $\tilde{U}_r$ on $\tilde{Y}_r$ and the direct effect of $\bar{U}_t$ on $\bar{Y}_t$. However, note that $\beta_Y$ was used to denote the direct effect of $C_r$ on $\tilde{Y}_r$, while $\omega_Y$ was used to denote the direct effect of $C$ on $\bar{Y}_t$. This emphasizes that step (b) allows for the direct effects of observed covariate on the outcome assumed in the power calculations to not be equal the corresponding effects in the data-generating model. This allows assessing the potential influence of misspecifying these effects in the power calculation, which is important because both approaches used to determine these effects in the power calculations have limitations, as described above. In the simulations, we assessed the following strategies for determining $\beta_Y$:

(i) Set $\beta_Y = \omega_Y$ – i.e., assume true coefficients are known (strategy 1).

(ii) Set $\beta_Y = 0$ – ie., assume no covariate has a direct effect on the outcome (strategy 2).

(iii) Fit a linear regression model for $\bar{Y}_t$ given $\bar{X}_t$ and $C$, and use the resulting regression coefficients for the covariates as $\beta_Y$ (strategy 3.1).

(iv) Apply backwise selection to (iii) above, removing the covariate with the largest P-value until doing so reduced some Information Criterion (IC) of the model. After filtering, the coefficients of the covariates that remained in the model were assigned to the corresponding positions in $\beta_Y$; for others, the corresponding positions in $\beta_Y$ were set to zero. The following ICs were considered: Akaike IC (AIC), corrected AIC (AICc), Hannan-Quinn IC (HQIC) and BIC (strategies 3.2-3.5, respectively). These IC are sorted from the least to the most stringent.

(v) Same as (iii), but the model does not include $\bar{X}_t$ (strategy 4.1).

(vi) Same as (iv) but applied to (v) (strategies 4.2-4.5).

For the simulations assessing the performance of the power calculation method, we considered two main scenarios (see LINK TO EXCEL IN GITHUB for a full description of all parameters and results). In scenario 1, no covariate had a direct effect on $\bar{Y}_t$ – i.e., $c_Y = 0$. Moreover, the parameters $\gamma$ and $\delta$ and were varied as follows: $\gamma \in \left\{\frac{|\delta|}{2\sqrt{0.9}}, \frac{1}{2}\left(\frac{|\delta|}{2\sqrt{0.9}} + \frac{0.9-p_U}{1-p_X}\right), \frac{0.9-p_U}{1-p_X}\right\}$ and $\delta \in \{0.4, 0.8\}$. This scenario also considered more relaxed and more stringent P-values for classifying a covariate as satisfying the conditions of the Theorem: P<0.3 (condition (i)) and P≥0.05 (condition (ii)) for the more relaxed classification; and P<0.05 (condition (i)) and P≥0.3 (condition (ii)) for the more stringent classification. In scenario 2, some covariates were allowed to have direct effects on the outcome – i.e., $c_Y \in \{3,6\}$. Again, different values of $\delta \in \{-0.6, 0.2, 0.6\}$ were considered, as well as different values for $\zeta \in \{-0.6, 0, 0.2, 0.6\}$.

Selected results of the assessment of the power calculation method are shown in Supplementary Tables 6 and 7. For, consider Supplementary Table 6. Let $k_i$ denote some combination of parameters. Under $k_i$, $T = 1000$ datasets were simulated, and to each dataset the confounding assessment method was applied (call this the "true result"). Then, the proportion of simulated datasets containing at least one covariate classified as satisfying both conditions of the Theorem was calculated. Moreover, for each simulated dataset, the power calculation method was applied, which simulated $R = 200$ for each one of the $T$ simulations. Using the set of $R$ simulated datasets, the proportion of simulated datasets containing at least one covariate classified as satisfying both conditions of the Theorem was calculated (this is an estimate of the true result, so can be called "estimated result"). These were then averaged over the $T$ simulations. Therefore, a positive (mean estimated result) – (true result) expresses a conservative result, in the sense that it suggests that the confounding test fails to detect unmeasured confounding more commonly than it actually does. For example: suppose that the true result is 1% and the estimated result is 30%. Then, the estimated result overestimates the likelihood of failing to detect unmeasured confounding, thus weakening interpreting a situation where both conditions of the Theorem are satisfied as a strong indication of unconfoundedness, which would wrongly appear to be quite common even in the presence of unmeasured confounding.

Supplementary Table 6 shows descriptive statistics for (mean estimated result) – (observed result), expressed in percentage point differences, over all combinations of parameters under a given scenario. For example, for strategy 1 (which assumes direct effect parameters are known) under scenario 1 (where $c_Y = 0$), the maximum was 4.9. This means that, over all combinations of parameters under this scenario, the largest positive difference between mean estimated and observed result was 4.9 percentage points. For strategy 1, the method was slightly conservative under scenario 1 and somewhat more conservative under scenario 2. Strategy 2 is the same as strategy 1 under scenario 1. Under scenario 2, this strategy is overly conservative, which is expected since assuming all covariates have no direct effect on the outcome tends to mitigate statistical evidence for conditional association. However, strategies using estimated direct effect coefficients (3.1 and 4.1) were considerably non-conservative. This is expected since, given sampling error, the estimated direct effect coefficients often deviate from zero even if the true coefficients are zero, thus implying that the simulations performed under the power calculations assume non-zero direct effect coefficients for covariates that actually have no direct effects. However, this was mitigated by statistical filtering and, when filtering based on the BIC (strategies 3.5 and 4.5), the performance was comparable to the performance observed for strategy 1, which assumes known direct effect coefficients. Importantly, there were still some differences between strategies 3.5 and 4.5, thus reinforcing the importance of performing both in practice since the true direct effects will often be unknown. Similar trends were observed when considered at least two (instead of one) covariates classified as satisfying both conditions of the Theorem (Supplementary Table 7).

## 8. Supplementary discussion

In the main text, we argued that assuming the covariate adjustment set if minimally sufficient will often be implausible in practice. Nevertheless, if no assumption on the minimality of $C$ is warranted and all selected covariates that are associated with the exposure (conditional on the other covariates) are also associated with the outcome (given the exposure and the other covariates), one possibility is to expand the covariate set (as long the requirements outlined in the assumptions are satisfied) to include additional (and possibly redundant) variables. If the initial set of covariates was indeed minimally sufficient and the newly added covariate is exogenous (see the end of supplementary section 3 for a note on exogenous covariates), then, given our assumptions, such a newly added covariate would be expected to not be associated with the outcome given the exposure and the initial set of covariates. Of course, if faithfulness is violated or statistical power is low, such lack of association could happen even if the initial set was invalid. If such a newly added covariate is conditionally associated with the outcome, this would indicate the original adjustment set is invalid. Importantly, in this case the new covariate should not necessarily be added to the adjustment set. For example, if it is a valid instrument (conditional on the original adjustment set), then its inclusion could lead to bias amplification (i.e., larger bias due to conditioning on an instrument if the adjustment set is invalid).[8,9]

In addition to Entner and colleagues' method (described in detail supplementary section 5), other covariate selection approaches are worth describing. One of them is the so-called disjunctive cause criterion, which postulates that, if the entire set of measured covariates contains a subset that is a valid adjustment set, it is possible to select a valid adjustment set by selecting covariates that cause $X$ or $Y$ (or both), except those that are known to be instruments (to avoid bias amplification);[8,9] and proxies of unmeasured common causes of $X$ or $Y$.[10,11] This approach for covariate selection allows using partial causal knowledge (i.e., whether a covariate is a cause of $X$ and/or $Y$, with no requirement regarding the causal relationships between covariates themselves) to select a valid adjustment set. An important limitation of this approach is the assumption that there is a valid adjustment set within the set of measured covariates, which may be difficult to justify, although in general it is likely more plausible if the entire set of covariates is large, covers many different domains (socioeconomic, behavioural, genetic etc.) and accounts for a large proportion of the variance of $X$. If pre-specifying the set of covariates is undesirable (for example, when deciding which covariates to measure at the study design stage), a recently proposed method uses iterative graph expansion to learn from the user, who inputs causal assumptions bit by bit until a valid adjustment set is found (or the method declares that none exists).[12] In addition to being dynamic (in the sense that it interactively learns from the user), this method is efficient in the sense that it requires the user to provide only as many causal assumptions are required to find a valid adjustment set.

It should be noted that both the disjunctive cause criterion and the iterative graph expansion methods differs from the main application of our test, which aims at assessing whether a given set of covariates (selected using these criteria or any other) is a valid adjustment set. Indeed, the two approaches could be used in combination, with unconfoundedness assessment (which assesses if a selected covariate set is a valid adjustment set, but provides no guidance on how to select covariates apart from the assumptions the method requires) following covariate selection (which guides how to define the candidate adjustment set). Of note, even though it is sensible to not use known instruments for bias adjustment, such instruments could nevertheless be used for unconfoundedness assessment (as described above) of the selected adjustment set.

## 10.Supplementary Tables

**Supplementary Table 1. Simulation scenario 1: power of the test to detect unmeasured confounding according to sample size, unconditional association between the testing covariate and the exposure ($\beta_{C_1 \to X}$) and strength of unmeasured confounding ($\beta_U$).**

| Sample size | $\beta_{C_1 \to X}$ | $\beta_U$ | Rejection rate (%)[a] | | |
|---|---|---|---|---|---|
| | | | P<5% | P<1% | P<0.1% |
| 500 | 0.3 | 0.3 | 11.2% | 3.1% | 0.6% |
| 1000 | 0.3 | 0.3 | 18.9% | 6.9% | 1.3% |
| 5000 | 0.3 | 0.3 | 67.0% | 42.8% | 18.9% |
| 20000 | 0.3 | 0.3 | 99.8% | 98.6% | 92.4% |
| 5000 | 0.1 | 0.3 | 10.8% | 3.1% | 0.5% |
| 5000 | 0.2 | 0.3 | 32.1% | 13.1% | 3.8% |
| 5000 | 0.3 | 0.3 | 66.3% | 43.0% | 18.1% |
| 5000 | 0.4 | 0.3 | 92.1% | 80.3% | 55.2% |
| 5000 | 0.5 | 0.3 | 99.5% | 98.5% | 92.6% |
| 5000 | 0.3 | 0 | 5.0% | 0.9% | 0.1% |
| 5000 | 0.3 | 0.1 | 5.8% | 1.5% | 0.2% |
| 5000 | 0.3 | 0.2 | 17.5% | 6.3% | 1.3% |
| 5000 | 0.3 | 0.3 | 66.7% | 41.3% | 18.2% |
| 5000 | 0.3 | 0.4 | 99.2% | 95.9% | 84.8% |
| 5000 | 0.3 | 0.5 | 100.0% | 100.0% | 100.0% |

[a]Different P-values thresholds were used to infer whether $C_1$ and $Y$ are conditionally independent given $X$ and $C_2$.

**Supplementary Table 2. Simulation scenario 2: performance of the proposed test in the presence of residual confounding and model misspecification. For each scenario, 5000 datasets (each containing 20,000 observations) were simulated.**

| Model[A] | Specification[B] | Quantity | Scenario 2a | Scenario 2b |
|---|---|---|---|---|
| 1 | Linear | Rejection rate[C] | 3.2% | 100.0% |
| 1 | Quadratic | Rejection rate[C] | 100.0% | 100.0% |
| 2 | Linear | Mean estimated regression coefficient[D] | 0.000 | 0.461 |
| 2 | Linear | Rejection rate[C] | 5.0% | 5.0% |
| 2 | Quadratic | Mean estimated regression coefficient[D] | 0.000 | 0.453 |
| 2 | Quadratic | Rejection rate[C] | 5.0% | 100.0% |

A: Linear regression model for $X$ given covariates 1 and 2 (model 1) or for $Y$ given $X$ and the two covariates (model 2).
B: Linear: all dependent variables modelled as linear terms. Quadratic: covariate 2 modelled as a quadratic term.
C: Proportion of simulated datasets where the P-value for a T test of the null hypothesis that the coefficient for covariate 1 is zero was <5%.
D: Average value (over simulated datasets) of the regression coefficient for $X$ from model 2.

**Supplementary Table 3. Simulation scenario 3: performance of the test according to stringency of the statistical cutoffs to classify whether a covariate satisfies conditions (i) and (ii) of the Theorem, Pearson correlation between each pair of covariates ($\rho$), number of covariates that directly affect $X$ but not $Y$ ($c_X$), number of covariates that directly affect $Y$ but not $X$ ($c_Y$), proportion of variance of $X$ collectively explained by confounders in a correctly-specified model ($\theta_X$), proportion of variance of $Y$ collectively explained by confounders in a correctly-specified model ($\theta_Y$), data-generating models and model specification. For each combination of parameters, 5000 datasets (each containing 5000 observations) were simulated.**

| Scenario | Statistical cutoff stringency[A] | $\rho$[B] | $c_X$ | $c_Y$ | $\theta_X$[C] | $\theta_Y$[D] | True data-generating model[E] | | Model specification[F] | Mean bias[G] | Mean # covariates satisfying (i) | % of simulated datasets with L covariates satisfying (i) and (ii) | | | |
|---|---|---|---|---|---|---|---|---|---|---|---|---|---|---|---|
| | | | | | | | $f_X$ | $f_Y$ | | | | L=0 | L≥1 | L≥2 | L≥3 |
| 3a | Higher | 0 | 0 | 4 | 0.3 | 0.3 | Linear | Linear | 1 (Correct) | 0.000 | 14.2 | 99.9 | 0.1 | 0.0 | 0.0 |
| 3a | Higher | 0 | 4 | 4 | 0.3 | 0.3 | Linear | Linear | 1 (Correct) | 0.000 | 17.5 | 14.3 | 85.7 | 52.7 | 20.0 |
| 3a | Higher | 0 | 0 | 4 | 0.3 | 0.3 | Linear | Linear | 2 (Incorrect) | 0.197 | 10.7 | 88.4 | 11.6 | 0.8 | 0.0 |
| 3a | Higher | 0 | 4 | 4 | 0.3 | 0.3 | Linear | Linear | 2 (Incorrect) | 0.207 | 14.4 | 26.6 | 73.4 | 33.6 | 9.0 |
| 3a | Higher | 0 | 0 | 4 | 0.3 | 0.3 | Linear | Linear | 3 (Incorrect) | 0.241 | 5.3 | 89.0 | 11.0 | 0.7 | 0.0 |
| 3a | Higher | 0 | 4 | 4 | 0.3 | 0.3 | Linear | Linear | 3 (Incorrect) | 0.252 | 9.0 | 38.0 | 62.0 | 21.4 | 4.8 |
| 3b | Lower | 0 | 4 | 4 | 0.3 | 0.3 | Linear | Linear | 1 (Correct) | 0.000 | 17.6 | 5.6 | 94.4 | 75.4 | 40.6 |
| 3b | Lower | 0 | 4 | 4 | 0.3 | 0.3 | Linear | Linear | 3 (Incorrect) | 0.252 | 9.7 | 4.3 | 95.7 | 77.9 | 48.2 |
| 3c | Higher | 0 | 4 | 4 | 0.3 | 0.3 | Linear | Exponential | 3 (Incorrect) | 0.231 | 9.0 | 16.4 | 83.6 | 45.4 | 14.7 |
| 3c | Higher | 0 | 4 | 4 | 0.3 | 0.3 | Linear | Logarithmic | 3 (Incorrect) | 0.251 | 9.0 | 41.3 | 58.7 | 19.6 | 4.1 |
| 3c | Higher | 0 | 4 | 4 | 0.3 | 0.3 | Exponential | Exponential | 3 (Incorrect) | 0.253 | 8.8 | 22.2 | 77.8 | 37.5 | 10.8 |
| 3c | Higher | 0 | 4 | 4 | 0.3 | 0.3 | Exponential | Logarithmic | 3 (Incorrect) | 0.220 | 8.8 | 35.8 | 64.2 | 24.6 | 5.4 |
| 3c | Higher | 0 | 4 | 4 | 0.3 | 0.3 | Logarithmic | Exponential | 3 (Incorrect) | 0.220 | 8.9 | 16.0 | 84.0 | 48.2 | 16.0 |
| 3c | Higher | 0 | 4 | 4 | 0.3 | 0.3 | Logarithmic | Logarithmic | 3 (Incorrect) | 0.252 | 8.9 | 41.7 | 58.3 | 18.8 | 3.4 |
| 3d | Higher | 0 | 4 | 4 | 0.2 | 0.2 | Linear | Linear | 3 (Incorrect) | 0.162 | 7.8 | 12.9 | 87.1 | 54.1 | 19.9 |
| 3d | Higher | 0 | 4 | 4 | 0.2 | 0.6 | Linear | Linear | 3 (Incorrect) | 0.281 | 7.8 | 50.9 | 49.1 | 12.1 | 1.6 |
| 3d | Higher | 0 | 4 | 4 | 0.6 | 0.2 | Linear | Linear | 3 (Incorrect) | 0.327 | 9.9 | 19.3 | 80.7 | 45.1 | 16.0 |
| 3d | Higher | 0 | 4 | 4 | 0.6 | 0.6 | Linear | Linear | 3 (Incorrect) | 0.567 | 9.9 | 53.2 | 46.8 | 12.7 | 1.4 |
| 3e | Higher | 0.3 | 4 | 4 | 0.3 | 0.3 | Linear | Linear | 3 (Incorrect) | 0.321 | 7.9 | 41.1 | 58.9 | 21.7 | 5.9 |
| 3e | Higher | 0.6 | 4 | 4 | 0.3 | 0.3 | Linear | Linear | 3 (Incorrect) | 0.440 | 4.6 | 48.8 | 51.2 | 14.5 | 2.8 |
| 3f | Higher | 0 | 4 | 0 | 0.3 | 0.3 | Linear | Linear | 3 (Incorrect) | 0.252 | 9.0 | 36.2 | 63.8 | 24.9 | 5.4 |

A: Higher: a covariate satisfied condition (i) if its Bonferroni-corrected P-value was <5% (lower) or <0.1% (higher) and condition (ii) if its uncorrected P-value was ≥5% (lower) or ≥30% (higher).
B: True for pair of continuous covariates, but only approximate for pairs of binary covariates or a pair of one continuous and one binary variable.
C: $\theta_X$ is the proportion of the variance of $X$ collectively explained by confounders in a correctly-specified model. $\theta_X(c_X \div 12)$ is the proportion of variance of $X$ collectively explained by covariates that influence $X$, but not $Y$, in a correctly-specified model.
D: $\theta_Y$ is the proportion of the variance of $Y$ collectively explained by confounders in a correctly-specified model. $\theta_Y(c_Y \div 12)$ is the proportion of variance of $Y$ collectively explained by covariates that influence $Y$, but not $X$, in a correctly-specified model.
E: See section 5 in the Supplement for details.
F: This refers to assumed models for $X$ and $Y$, which do not necessarily correspond to the corresponding true data-generating models. Model specifications 1-3 are described in section 5 in the Supplement. Whether the assumed specification is correct or not is indicated in parenthesis.
G: Linear regression coefficient of $Y$ on $X$ given covariates under the corresponding model specification.

**Supplementary Table 4. Log-likelihood for different model specifications for the exposure $X$ and the outcome $Y$. Estimates were averaged over 50 iterations of 5-fold cross validation.**

| Model for $X$[A] | Log-likelihood | Model for $Y$[A] | Log-likelihood |
|---|---|---|---|
| 1.1 | -345.9 | 2.1 | -787.9 |
| 1.2 | -352.4 | 2.2 | -795.2 |
| 1.3 | -360.3 | 2.3 | -788.4 |
| 1.4 | -356.8 | 2.4 | -809.1 |
| 1.5 | -359.3 | 2.5 | -813.4 |
| 1.6 | -351.3 | 2.6 | -809.1 |
| - | - | 2.7 | -813.4 |

A: See supplementary section 6 for details.

**Supplementary Table 5. Results of power calculation under different parameters. Estimates show the proportion (out of 1000) of datasets simulated under the given parameters that had, among the covariates that satisfied condition (i) of the Theorem in the real data, at least one covariate that also satisfied condition (ii).**

| $\varphi$ | $\delta$ | Estimated probability (%) of at least one covariate satisfying conditions (i) and (ii) of the Theorem | |
|---|---|---|---|
| | | **A** | **B** |
| 0.1 | 0.2 | 63.5 | 62.6 |
| 0.1 | 0.4 | 53.1 | 53.9 |
| 0.2 | 0.2 | 61.3 | 61.2 |
| 0.2 | 0.4 | 45.8 | 45.9 |
| 0.4 | 0.2 | 60.5 | 61.1 |
| 0.4 | 0.4 | 41.6 | 44.0 |

$\varphi$: proportion of the variance of the simulated outcome explained by the direct effect of the simulated unmeasured confounder on the simulated outcome.
$\delta$: bias of the coefficient of a linear regression of simulated outcome on simulated exposure (as an estimator of the causal effect of simulated exposure on simulated outcome) due to unmeasured confounding.
A: Association between covariates and simulated outcome ($\beta_Y$) using the coefficients from a linear regression model of $Y$ given $X$ and $C$ followed by Bayesian Information Criterion (BIC)-based filtering.
B: Association between covariates and simulated outcome ($\beta_Y$) using the coefficients from a linear regression model of $Y$ given $C$ followed by BIC-based filtering.

**Supplementary Table 6. Summarized results of the assessment of the power calculation method. For each strategy for determining $\boldsymbol{\beta_Y}$ for the power calculations and each combination of parameters, the proportion of simulated datasets where at least one covariate was classified as satisfying both conditions (i) and (ii) of the Theorem was subtracted from the mean value (over all simulated datasets for that combination of parameters) of the corresponding proportions estimated by the power calculation method. Then, for each strategy for determining $\boldsymbol{\beta_Y}$, minimum, maximum, mean and median of such percentage point differences were taken over all combinations of parameters considered in each scenario. Positive numbers indicate that the power calculation method tends to overestimate the probability of failure to detect unmeasured confounding. Negative numbers indicate otherwise.**

| Strategy for | Scenario 1 | | | | Scenario 2 | | | |
|---|---|---|---|---|---|---|---|---|
| determining $\boldsymbol{\beta_Y}$ [A] | Minimum | Maximum | Mean | Median | Minimum | Maximum | Mean | Median |
| 1 | 0.0 | 4.9 | 0.9 | 0.0 | -0.7 | 17.9 | 5.4 | 3.4 |
| 2 | 0.0 | 4.9 | 0.9 | 0.0 | 1.6 | 47.9 | 27.9 | 33.3 |
| 3.1 | -42.4 | -0.2 | -14.9 | -9.5 | -36.6 | -12.9 | -21.6 | -20.4 |
| 3.2 | -20.3 | 0.0 | -5.4 | -2.2 | -18.1 | -2.0 | -7.8 | -6.5 |
| 3.3 | -20.1 | 0.0 | -5.3 | -2.1 | -17.9 | -1.8 | -7.7 | -6.4 |
| 3.4 | -9.1 | 0.1 | -1.7 | -0.5 | -5.7 | 9.0 | 0.6 | -0.3 |
| 3.5 | -3.4 | 0.2 | -0.3 | -0.1 | -0.7 | 17.5 | 6.0 | 5.6 |
| 4.1 | -5.6 | 0.1 | -1.7 | -0.4 | -36.6 | 26.1 | -5.3 | -9.5 |
| 4.2 | -0.9 | 3.7 | 0.4 | 0.0 | -18.3 | 20.6 | 0.1 | -2.8 |
| 4.3 | -0.9 | 3.7 | 0.4 | 0.0 | -18.2 | 20.5 | 0.2 | -2.8 |
| 4.4 | -0.1 | 4.7 | 0.9 | 0.0 | -6.2 | 18.9 | 3.4 | -0.4 |
| 4.5 | 0.0 | 4.8 | 0.9 | 0.0 | -0.6 | 19.9 | 6.1 | 3.6 |

A: See supplementary section 7 for details.

**Supplementary Table 7. Summarized results of the assessment of the power calculation method. For each strategy for determining $\boldsymbol{\beta_Y}$ for the power calculations and each combination of parameters, the proportion of simulated datasets where at least two covariates was classified as satisfying both conditions (i) and (ii) of the Theorem was subtracted from the mean value (over all simulated datasets for that combination of parameters) of the corresponding proportions estimated by the power calculation method. Then, for each strategy for determining $\boldsymbol{\beta_Y}$, minimum, maximum, mean and median of such percentage point differences were taken over all combinations of parameters considered in each scenario. Positive numbers indicate that the power calculation method tends to overestimate the probability of failure to detect unmeasured confounding. Negative numbers indicate otherwise.**

| Strategy for | Scenario 1 | | | | Scenario 2 | | | |
|---|---|---|---|---|---|---|---|---|
| determining $\boldsymbol{\beta_Y}$[A] | Minimum | Maximum | Mean | Median | Minimum | Maximum | Mean | Median |
| 1 | -1.6 | 12.1 | 2.4 | 0.0 | -1.3 | 11.6 | 3.9 | 3.0 |
| 2 | -1.6 | 12.1 | 2.4 | 0.0 | 12.2 | 70 | 55.7 | 61.3 |
| 3.1 | -45.4 | -2.4 | -28.4 | -33.3 | -33.0 | -4.1 | -15.7 | -14.3 |
| 3.2 | -28.8 | -0.3 | -12.1 | -11.2 | -9.9 | 0.8 | -5.5 | -6.0 |
| 3.3 | -28.6 | -0.3 | -12.1 | -11 | -9.8 | 0.9 | -5.4 | -5.8 |
| 3.4 | -14.6 | -0.1 | -4.2 | -3.4 | -3.3 | 7.8 | 1.6 | 1.3 |
| 3.5 | -5.0 | 2.9 | -0.7 | -0.3 | -0.8 | 14.2 | 6.9 | 6.8 |
| 4.1 | -19.7 | -0.2 | -6.0 | -1.7 | -26.2 | 17.3 | -7.0 | -13.3 |
| 4.2 | -4.7 | 8.9 | 0.7 | -0.1 | -18.0 | 15.6 | -1.1 | -4.4 |
| 4.3 | -4.7 | 9.0 | 0.7 | -0.1 | -17.9 | 15.5 | -1.1 | -4.4 |
| 4.4 | -2.2 | 11.4 | 2.1 | 0.0 | -10.5 | 15.8 | 2.5 | 1.3 |
| 4.5 | -1.8 | 11.8 | 2.3 | 0.0 | -2.5 | 17.8 | 6.0 | 6.4 |

A: See supplementary section 7 for details.

**11.Supplementary Figure**

**Supplementary Figure 1. Directed acyclic graphs illustrating scenarios with measurement error in one or more covariates. $V'$ represents the measured version of $V$, and sources of measurement error are represented by $E_V$. $Z$ and $D$ denote measured covariates, $M$ denotes the error-free version of $D$, $U$ denotes an unmeasured variable, $X$ and $Y$ denote measured exposure and outcome, while $T$ and $G$ denote their error-free counterparts. Dashed and gray lines indicate effects of special interest in each scenario, as discussed in the text.**

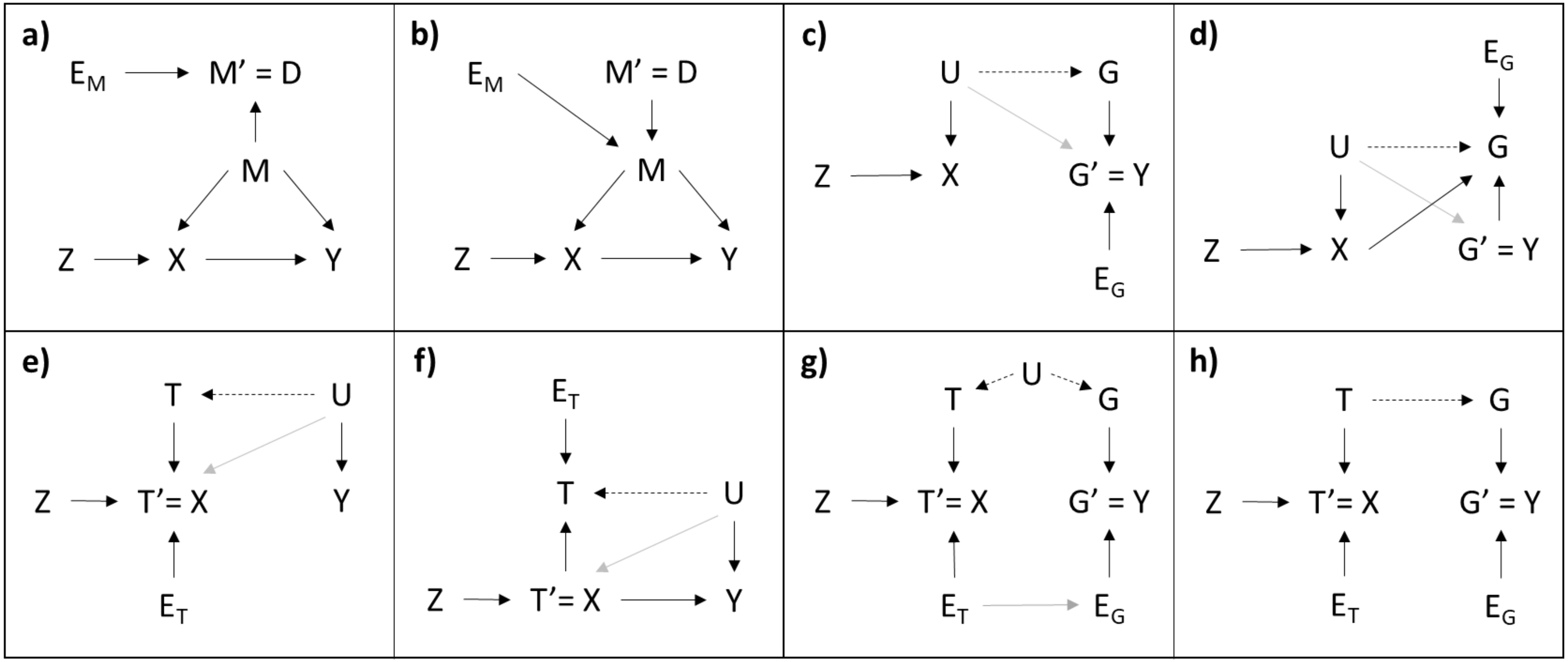